\documentclass[10pt]{article}
\date{}  
\usepackage{hyperref}
\usepackage{amssymb, amsmath}

\usepackage{amsmath}
\usepackage{latexsym}
\usepackage{amssymb}

\usepackage{amsthm}

\font\tengoth=eufm10 at 10pt
\font\sevengoth=eufm7 at 6pt
\newfam\gothfam
\textfont\gothfam=\tengoth
\scriptfont\gothfam=\sevengoth

\newcommand{\mlabel}{\label}

\newcommand{\fB}{{\mathfrak B}}

\newcommand{\g}{{\mathfrak g}}

\newcommand{\fh}{{\mathfrak h}}

\newcommand{\fq}{{\mathfrak q}}

\renewcommand{\:}{\colon}
\newcommand{\1}{\mathbf{1}}

\newcommand{\cA}{\mathcal{A}}

\newcommand{\cE}{\mathcal{E}}

\newcommand{\cH}{\mathcal{H}}

\newcommand{\cN}{\mathcal{N}}

\newcommand{\cS}{\mathcal{S}}

\newcommand\bx{{\bf{x}}}

\newcommand{\dd}{{\tt d}}

\newcommand{\subeq}{\subseteq}
\newcommand{\supeq}{\supseteq}

\newcommand{\eps}{\varepsilon}

\newcommand{\shalf}{{\textstyle{\frac{1}{2}}}}

\newcommand{\N}{{\mathbb N}}

\newcommand{\R}{{\mathbb R}}
\newcommand{\C}{{\mathbb C}}

\newcommand{\K}{{\mathbb K}}

\newcommand{\bP}{{\mathbb P}}

\newcommand{\bE}{{\mathbb E}}

\newcommand{\bS}{{\mathbb S}}

\renewcommand{\hat}{\widehat}

\renewcommand{\tilde}{\widetilde}

\newcommand{\Aff}{\mathop{{\rm Aff}}\nolimits}

\newcommand{\GL}{\mathop{{\rm GL}}\nolimits}
\newcommand{\SL}{\mathop{{\rm SL}}\nolimits}

\newcommand{\PGL}{\mathop{{\rm PGL}}\nolimits}

\newcommand{\PSL}{\mathop{{\rm PSL}}\nolimits}
\newcommand{\SO}{\mathop{{\rm SO}}\nolimits}

\newcommand{\OO}{\mathop{\rm O{}}\nolimits}

\newcommand{\U}{\mathop{\rm U{}}\nolimits}

\renewcommand{\Re}{\mathop{{\rm Re}}\nolimits}

\newcommand{\Conf}{\mathop{\rm Conf{}}\nolimits}

\newcommand{\Diff}{\mathop{{\rm Diff}}\nolimits}

\newcommand{\id}{\mathop{{\rm id}}\nolimits}

\renewcommand{\dim}{\mathop{{\rm dim}}\nolimits}

\newcommand{\sgn}{\mathop{{\rm sgn}}\nolimits}

\newcommand{\Spann}{\mathop{{\rm span}}\nolimits}
\newcommand{\ev}{\mathop{{\rm ev}}\nolimits}

\newcommand{\vphi}{\varphi}
\renewcommand{\phi}{\varphi}

\newcommand{\Rarrow}{\Rightarrow}
 
\newcommand{\oline}{\overline}

\newcommand{\la}{\langle}
\newcommand{\ra}{\rangle}

\newcommand{\Mot}{{\rm Mot}}

\newcommand{\res}{\vert}

\newcommand{\Spec}{{\rm Spec}}

\newcommand{\ssssarr}{\hbox to 15pt{\rightarrowfill}}
\newcommand{\sssarr}{\hbox to 20pt{\rightarrowfill}}
\newcommand{\ssarr}{\hbox to 30pt{\rightarrowfill}}
\newcommand{\sarr}{\hbox to 40pt{\rightarrowfill}}
\newcommand{\arr}{\hbox to 60pt{\rightarrowfill}}
\newcommand{\larr}{\hbox to 60pt{\leftarrowfill}}
\newcommand{\Arr}{\hbox to 80pt{\rightarrowfill}}

\def\theoremname{Theorem}
\def\propositionname{Proposition}
\def\corollaryname{Corollary}
\def\lemmaname{Lemma}
\def\remarkname{Remark}
\def\conjecturename{Conjecture} 

\def\definitionname{Definition}
\def\exercisename{Exercise}
\def\examplename{Example}
\def\examplesname{Examples}
\def\problemname{Problem}
\def\problemsname{Problems}

\def\satzname{Satz} 
\def\koroname{Korollar}
\def\folgname{Folgerung}
\def\bemerkname{Bemerkung}
\def\aufgname{Aufgabe}

\def\beisname{Beispiel}
\def\beissname{Beispiele}
\def\bewname{Beweis}

\def\@thmcounter#1{\noexpand\arabic{#1}}
\def\@thmcountersep{}
\def\@begintheorem#1#2{\it \trivlist \item[\hskip 
\labelsep{\bf #1\ #2.\quad}]}
\def\@opargbegintheorem#1#2#3{\it \trivlist
      \item[\hskip \labelsep{\bf #1\ #2.\quad{\rm #3}}]}
\makeatother
\newtheorem{theor}{\theoremname}[section]
\newtheorem{propo}[theor]{\propositionname}
\newtheorem{coro}[theor]{\corollaryname}
\newtheorem{lemm}[theor]{\lemmaname}

\newenvironment{thm}{\begin{theor}\it}{\end{theor}}

\newenvironment{prop}{\begin{propo}\it}{\end{propo}}

\newenvironment{cor}{\begin{coro}\it}{\end{coro}}

\newenvironment{lem}{\begin{lemm}\it}{\end{lemm}}

\newtheorem{rema}[theor]{\remarkname}

\newenvironment{remark}{\begin{rema}\rm}{\end{rema}}
\newenvironment{rem}{\begin{rema}\rm}{\end{rema}}

\newtheorem{stepnow}[theor]{}

\newtheorem{defin}[theor]{\definitionname} 

\newenvironment{definition}{\begin{defin}\rm}{\end{defin}}
\newenvironment{defn}{\begin{defin}\rm}{\end{defin}}

\newtheorem{exerc}{\exercisename}[section]

\newtheorem{exa}[theor]{\examplename}

\newenvironment{example}{\begin{exa}\rm}{\end{exa}}
\newenvironment{ex}{\begin{exa}\rm}{\end{exa}}

\newtheorem{exas}[theor]{\examplesname}

\newenvironment{exs}{\begin{exas}\rm}{\end{exas}}

\newtheorem{conj}[theor]{\conjecturename}

\newtheorem{pro}[theor]{\problemname}

\newenvironment{prob}{\begin{pro}\rm}{\end{pro}}

\newtheorem{prs}[theor]{\problemsname}

\newtheorem{aufg}{\aufgname}[section]

\newenvironment{prf}{\begin{proof}}{\end{proof}}
 
\newcommand{\pmat}[1]{\begin{pmatrix} #1 \end{pmatrix}}

{\hfill\qed\end{trivlist}}

\newenvironment{beweis*}{\begin{trivlist}\item[\hskip%
\labelsep{\bf\bewname.\quad}]}%
{\end{trivlist}}

\newtheorem{satzn}[theor]{\satzname}

\newtheorem{koro}[theor]{\koroname}

\newtheorem{folg}[theor]{\folgname}

\newtheorem{bem}[theor]{\bemerkname}

\newtheorem{aufgn}[theor]{\aufgname}

\newtheorem{beis}[theor]{\beisname}

\newtheorem{beiss}[theor]{\beissname}

\newcommand{\CR}{\mathop{{\rm CR}}\nolimits}

\newcommand{\E}{{\bE}}

\begin{document}

\title{Reflection negative kernels and fractional Brownian motion}

\author{P. Jorgensen, K.-H. Neeb, G. \'Olafsson} 

\maketitle

\begin{abstract} 
In this article we study the connection of fractional Brownian motion, representation theory and
reflection positivity in quantum physics. We introduce and study reflection
positivity for affine isometric actions of a Lie group on
a Hilbert space $\cE$ and
show in particular 
that fractional Brownian motion for Hurst index $0<H\le 1/2$ is reflection
positive and leads via reflection positivity to an infinite dimensional Hilbert space if
$0<H <1/2$. We also study projective invariance of fractional Brownian motion and
relate this to the complementary series representations of $\mathrm{GL}_2(\R )$. 
 We relate this to a measure preserving action on a Gaussian $L^2$-Hilbert space
$L^2(\cE)$. 
\end{abstract}

\tableofcontents 
\section{Introduction} 
\mlabel{sec:1}

In this paper we continue our investigations of the representation theoretic 
aspects of \textit{reflection positivity} and its relations to 
stochastic processes (\cite{JNO16a, JNO16b}). 
This is a basic concept in constructive quantum 
field theory \cite{GJ81, Kl77, KL82, JO00}, 
where it arises as a requirement on the euclidean side to establish a 
duality between euclidean and relativistic quantum field theories \cite{OS73}. 
It is closely related to ``Wick rotations'' or 
``analytic  continuation'' in the time variable 
from the real  to the imaginary axis. 

The underlying structure is that of a 
{\it reflection positive Hilbert space}, introduced in \cite{NO14}. 
This is a triple $(\cE,\cE_+,\theta)$, 
where $\cE$ is a Hilbert space, $\theta : \cE \to \cE$ is a unitary involution
and $\cE_+$ is a closed subspace of $\cE$ which is $\theta$-positive in the sense that 
the hermitian form $\langle u,\theta v\rangle$ is
positive semidefinite on $\cE_+$.
We write $\widehat\cE$ for the corresponding Hilbert space 
and $q \colon \cE_+ \to \widehat\cE, \xi \mapsto \hat\xi$ for the canonical map. 

To relate this to group representations, let us call a triple 
$(G,S,\tau)$ a {\it symmetric semigroup} if $G$ is a Lie group, 
$\tau$ is an involutive automorphism of $G$ and $S \subeq G$ a 
subsemigroup
invariant under the involution $s \mapsto s^\sharp := \tau(s)^{-1}$.  
The Lie algebra 
$\g$ of $G$ decomposes into $\tau$-eigenspaces $\g = \fh \oplus \fq$ 
and we obtain the {\it Cartan dual Lie algebra} $\g^c=\fh\oplus i\fq$. 
We write $G^c$ for a Lie group with Lie algebra $\g^c$. 
The prototypical pair $(G,G^c)$ consists of the euclidean 
motion group $E(d) = \R^d \rtimes \OO_d(\R)$ and the orthochronous 
Poincar\'e group $P(d)^\uparrow =  \R^d \rtimes \OO_{1,d-1}(\R)^\uparrow$. 
If $(G,H,\tau)$ is a symmetric Lie group and
$(\cE,\cE_+,\theta)$ a reflection positive Hilbert space, then we say that 
a unitary representation $U \colon G \to \U(\cE)$ is {\it reflection positive 
with respect to $(G,S,\tau)$} if 
\begin{equation}
  \label{eq:pi-tau-rel}
 U_{\tau(g)} = \theta U_g\theta \quad \mbox { for } g \in G 
\quad \mbox{ and } \quad U_S\cE_+\subeq\cE_+.
\end{equation}

If $(\pi,\cE)$ is a reflection positive representation of $G$ on 
$(\cE,\cE_+, \theta)$, then $\widehat U_s q(v) := q(U_s v)$ 
defines a representation
$(\widehat U, \widehat\cE)$ of the involutive semigroup $(S,\sharp)$ by contractions 
(\cite[Lemma 1.4]{NO14}, \cite{JO00} or \cite[Prop.~3.3.3]{NO18}). 
However, if $S$ has interior points, we would like to have a 
unitary representation $U^c$ of a Lie group $G^c$ with 
Lie algebra $\g^c$ on $\widehat\cE$ whose derived representation 
is compatible with the representation of~$S$. 
If such a representation exists, then we call 
$(U, \cE)$ a {\it euclidean realization} 
of the representation $(U^c,\widehat\cE)$ of $G^c$. 
Sufficient conditions for the existence of $U^c$ have been developed in 
\cite{MNO15}. 

Althought this is a rather general framework, 
the present paper is only concerned with very concrete aspect of reflection 
positivity. The main new aspect we introduce is a notion of 
reflection positivity for affine isometric actions of a symmetric semigroup 
$(G, S,\tau)$ on a real Hilbert space.  
Here $\cE_+$ is naturally defined by the closed subspace 
generated by the $S$-orbit of the origin. On the level 
of positive definite functions, this leads to the notion of a reflection 
negative function. For $(G,S,\tau) = (\R,\R_+,-\id_\R)$, 
reflection negative functions $\psi$ are easily determined because 
reflection negativity is equivalent to $\psi\res_{(0,\infty)}$ being a Bernstein 
function (\cite{JNO18}).  
An announcement of some of the results in the present paper appeared in~\cite{JNO16b}.

For a group $G$, affine isometric actions 
$\alpha_g \xi = U_g \xi + \beta_g$ on a real Hilbert space $\cE$ 
are encoded in real-valued negative definite functions $\psi(g) = \|\beta_g\|^2$ 
satisfying $\psi(e) = 0$ (cf.~\cite{Gu72, HV89}). Especially 
for $G = \R$, these structures have manifold applications in various 
fields of mathematics (see for instance \cite{Ka81}, \cite{Le80}, \cite{Mas72},  
and also \cite{Fu05} for the generalization to {\it spirals} which corresponds 
to actions of $\R$ by affine conformal maps). For the 
group $G = (\R,+)$, the homogeneous function 
$\psi(x) = |x|^{2H}$ is negative definite if and only if $0 \leq H \leq 1$, 
and this leads to the positive definite kernels 
\[ C^H(s,t) := \frac{1}{2}(|s|^{2H} + |t|^{2H} - |s-t|^{2H}), \]
which for $0 < H < 1$ are the covariance kernels of 
fractional Brownian motion with Hurst index $H$ 
(\cite{BOZ08}, \cite{AL08}, \cite{AJ12, AJ15}, \cite{AJP12}). 

One of the central results of this paper is an extension 
of the well-known projective invariance of Brownian 
motion in the sense of P.~L\'evy (cf.~\cite[\S I.2]{Le65} and 
\cite{HKNY69}) to fractional Brownian motion. 
Here we use the identification of $\R_\infty := \R \cup \{\infty\}$ 
with the real projective line, which leads to the action 
of $\GL_2(\R)$ by M\"obius transformations 
$g.x = \frac{ax + b}{cx+d}$ for $g = \pmat{a & b \\ c &d}$. 
Starting from a realization of fractional Brownian motion 
$(B^H_t)_{t \in \R}$ with Hurst index $H \in (0,1)$ in a 
suitable Hilbert space $\cH_H$ by the functions 
\begin{equation}
  \label{eq:bth}
b^H_t  := \sgn(t) \chi_{[t \wedge 0, t \vee 0]} 
= \chi_{[0,\infty)} - \chi_{[t,\infty)}, \qquad  t \in \R, 
\end{equation}
we associate to every pair of distinct points $\alpha, \gamma$ in $\R_\infty$ 
a normalized process whose covariance kernels 
$C^H_{\alpha,\gamma}$ transform naturally under M\"obius transformations in 
the sense that 
\begin{equation}
  \label{eq:ker-inv}
C^H_{g.\alpha, g.\beta}(g.s,g.t)  = C^H_{\alpha,\beta}(s,t).
\end{equation}
Here the normalized fractional Brownian motion 
$\tilde B^H_t = |t|^{-H} B^H_t$ has the covariance kernel $C^H_{0,\infty}$ 
and the transformed process $\tilde B^H_{g.t}$  
is equivalent to the original one. 

The structure of this paper is as follows. 
In Section~\ref{sec:2} we briefly recall the general background of 
reflection positive Hilbert spaces and representations 
and in Section~\ref{sec:4} we introduce reflection positive 
affine isometric actions $U \: G \to \Mot(\cE)$ on real Hilbert spaces~$\cE$. 
Since the group $\Mot(\cE)$ has a natural unitary representation 
on the Fock space $\Gamma(\cE)$, the $L^2$-space of the 
canonical Gaussian measure of $\cE$, affine isometric representations 
are closely linked with symmetries of Gaussian stochastic processes 
for which $G$ acts on the corresponding index set. 
This is made precise in Appendix~\ref{subsec:statinc}, where we 
discuss the measure preserving $G$-action corresponding 
to a stochasic process with stationary increments. For square 
integrable processes, this connects with 
affine isometric actions on Hilbert spaces. 

To pave the way for the analysis of the interaction of fractional 
Brownian motion with unitary representations, we 
introduce in Section~\ref{sec:5} a family of unitary 
representations $(U^H, \cH_H)_{0 < H < 1}$ of $\GL_2(\R)$, 
respectively its projective quotient $\PGL_2(\R)$, i.e., the 
group of M\"obius transformations on the real projective line. 
For $H =\shalf$ this is the natural representation on $L^2(\R)$ 
(belonging to the principal series), whereas for $H \not=\shalf$ 
it belongs to the complementary series (\cite{JO00}, \cite{vD09}). 
The Hilbert spaces $\cH_H$ are obtained from positive definite 
distribution kernels by completion of $\cS(\R)$ with respect to the scalar 
product 
\[  \la \xi,\eta\ra_H 
= -\frac{1}{2} \int_\R \int_\R \oline{\xi'(x)}\eta'(y) |x-y|^{2H}\, dx\, dy. \] 
In Section~\ref{sec:6} we realize fractional Brownian motion 
in a very natural way in terms of the cocycle \eqref{eq:bth} 
defining an affine isometric action of the translation group $(\R,+)$ on $\cH_H$. 
Acting with the group $\GL_2(\R)$ on these functions leads naturally 
to the projective invariance of fractional Brownian motion,
both on the level of the normalized kernels as in \eqref{eq:ker-inv}, 
and with respect to our concrete realization (Theorem~\ref{thm:f-ker-rel}). 

Reflection positivity is then explored in Section~\ref{sec:7}. 
For $\alpha\not=\gamma$ in $\R_\infty \cong \bS^1$ 
we consider a reflection $\theta$ with a fixed point and exchanging $\alpha$ and $\gamma$. 
Here our main result is Theorem~\ref{thm:6.3}, asserting that 
the normalized kernels $C^H_{\alpha,\gamma}$ on the complement 
of the two-element set $\{\alpha,\gamma\}$ in $\R_\infty \cong \bS^1$ 
is  reflection positive with respect to $\theta$ 
if and only if $H \leq \shalf$. 
In particular, this implies reflection positivity for a Brownian 
bridge on a real interval $[\alpha,\gamma]$ with respect to the reflection 
in the midpoint. Reflection positivity for the 
complementary series representations of $\SL_2(\R)$ has already been 
observed in \cite{JO00}, where the representation 
$U^c$ is identified as a holomorphic discrete series representation. 

Reflection positivity for the affine action of the translation 
group in $\cH_H$ defined by the cocycle $b^H_t$ realizing fractional 
Brownian motion is studied in Section~\ref{sec:8}. 
Although we always have involutions that lead to reflection 
positive Hilbert spaces in a natural way, 
only for $H \leq \shalf$ we obtain reflection positive 
affine actions of $(\R,\R_+,-\id)$. 
We conclude Section~\ref{sec:8} with a discussion of the increments 
of a $1$-cocycle $(\beta_t)_{t \in \R}$ defining an affine isometric 
action. In particular, we characterize cocycles with 
orthogonal increments as those corresponding to multiples of 
Brownian motion. Note that the increments of fractional 
Brownian motion are positively correlated for $H \geq \shalf$ and 
negatively correlated for $H \leq \shalf$. 
We conclude this paper with a brief discussion of some related 
results concerning higher dimensional spaces in Section~\ref{sec:9}. 
We plan to return to the corresponding representation theoretic 
aspects in the near future. 

In order not to distract the reader from the main line of the paper, 
we moved several auxiliary tools and some definitions and calculations 
into appendices: 
Appendix~\ref{sec:a.2} deals with affine isometries and 
positive definite kernels and Appendix~\ref{sec:a.1} reviews 
some properties of stochastic processes. In particular, we provide 
in Proposition~\ref{prop:a.4b} a representation theoretic 
proof for the L\'evy--Khintchine formula for the real line, 
which represents a negative definite function in terms of its 
spectral measure (\cite{AJ12, AJ15, AJL17}, \cite[Thm.~32]{Le65}). 
Appendix~\ref{app:2} briefly recalls the measure theoretic perspective 
on Second Quantization, Appendix~\ref{app:d} contains the verification 
that the representations $U^H$ mentioned above are unitary, and 
Appendix~\ref{app:e} contains a calculation of the spectral measure for 
fractional Brownian motion.

A different kind of projective invariance, in the path parameter $t$, 
for one-dimensional Brownian motion has been observed by 
S.~Takenaka in \cite{Ta88a}: For a Brownian motion $(B_t)_{t \in \R}$, the process 
\[ B^g_t := (ct + d) B_{g.t} - ct \cdot B_{g.\infty} - d \cdot B_{g.0}, \qquad 
g = \pmat{a & b \\ c &d} \in \SL_2(\R), \qquad t \in \R_\infty, 
g.t \not=\infty,\] 
also is a Brownian motion, and the relation $(B^g)^h = B^{gh}$ leads to a 
unitary representation of $\SL_2(\R)$ on the realization space. From that 
he derives the projective invariance in the sense of L\'evy, and he argues 
that his method does not extend to fractional Brownian motion. 
In \cite{Ta88b}, Takenaka shows that the representation of $\SL_2(\R)$ 
he obtains belongs to the discrete series, so that it is differet from ours. 
He also hints at the possibility of extending Hida's method 
\cite{HKNY69} to fractional Brownian motion, and in a certain 
sense this is carried out in the present paper.

\section{Reflection positive functions and representations} 
\mlabel{sec:2}

Since our discussion is based on positive definite kernels 
and the associated Hilbert spaces (\cite{Ar50}, \cite[Ch.~I]{Ne00}, \cite{NO18}), 
we first recall the pertinent definitions. 
As customary in physics, 
we follow the convention that the inner product of a complex
 Hilbert space is linear in the second argument.

\begin{defn}   \label{def:8.1.1} 
(a) Let $X$ be a set. A kernel $Q \colon X \times X \to \C$ is called 
{\it hermitian} if $Q(x,y) = \overline{Q(y,x)}$. 
A hermitian kernel $Q$ is called
 {\it positive definite} if for 
$x_1, \ldots, x_n \in X, c_1, \ldots, c_n \in \C$, we have 
$\sum_{j,k=1}^n c_j \overline{c_k} Q(x_j, x_k) \geq 0$.
It is called {\it negative definite} if 
$\sum_{j,k=1}^n c_j \overline{c_k} Q(x_j, x_k) \leq 0$ 
holds for $x_1, \ldots, x_n \in X$ and $c_1, \ldots, c_n \in \C$ 
with $\sum c_j=0$ (\cite{BCR84}). 

(b) If $(S,*)$ is an involutive semigroup, then $\vphi \colon S \to \C$ is 
called {\it positive (negative) definite} if the kernel $(\vphi(st^*))_{s,t\in S}$ 
is positive (negative) definite. 
If $G$ is a group, then we consider it as an involutive semigroup 
with $g^* := g^{-1}$ and definite positive/negative definite functions accordingly.  
\end{defn} 

We shall use the following lemma to translate between positive 
definite and negative definite kernels 
(\cite[Lemma~3.2.1]{BCR84}): 
\begin{lem}\mlabel{lem:bcr} 
Let $X$ be a set, $x_0 \in X$ and $Q \: X \times X \to \C$  be a hermitian kernel. 
Then the kernel 
\[ K(x,y) := Q(x,x_0) + Q(x_0,y) - Q(x,y) - Q(x_0, x_0) \] 
is positive definite if and only if $Q$ is negative definite. 
\end{lem}

\begin{remark} \mlabel{rem:schoen} According to Schoenberg's Theorem 
\cite[Thm.~3.2.2]{BCR84}, 
a kernel $Q \: X \times X \to \C$ is negative definite 
if and only if, for every $h > 0$, the kernel 
$e^{-h Q}$ is positive definite. 
\end{remark}

\begin{remark} \label{rem:kerspace} 
Let $X$ be a set, $K \colon X \times X \to \C$ be a positive definite 
kernel and $\cH_K \subeq \C^X$ be the corresponding {\it reproducing kernel 
Hilbert space}. This is the unique Hilbert subspace of $\C^X$ on which all 
point evaluations $f \mapsto f(x)$ are continuous and given by 
\[ f(x) = \la K_x, f \ra \quad \mbox{ for } \quad 
K(x,y) = K_y(x) = \la K_x, K_y \ra.\] 
Then the map $\gamma \colon X \to \cH_K, \gamma(x) = K_x$ has total range 
and satisfies $K(x,y) = \la \gamma(x),\gamma(y)\ra$. The latter property 
determines the pair $(\gamma, \cH_K)$ up to unitary equivalence 
(\cite[Ch.~I]{Ne00}). 
\end{remark}

\begin{defn} \mlabel{def:1.2}  A {\it reflection positive Hilbert space} is a triple 
$(\cE,\cE_+,\theta)$, where 
$\cE$ is a Hilbert space, $\theta$ a unitary involution and 
$\cE_+$ is a closed subspace which is {\it $\theta$-positive} in the sense that 
the hermitian form $\la \xi,\eta\ra_\theta := \la \xi, \theta \eta\ra$ is positive 
semidefinite on~$\cE_+$. 

For a reflection positive Hilbert space  $(\cE,\cE_+,\theta)$, 
let $\cN:=\{\xi \in\cE_+\: \la \xi, \theta \xi\ra =0\}$ and 
write $\hat\cE$ for the completion of $\cE_+/\cN$ with respect to the inner product   
$\la \cdot, \cdot \ra_\theta$. We write 
 $q \: \cE_+ \to \hat\cE, \xi \mapsto \hat\xi$ for the canonical map. 
\end{defn}

\begin{example} \mlabel{ex:1.3} 
Suppose that $K \colon X \times X \to \C$ is a positive definite kernel 
and $\tau \colon X \to X$ is an involution 
leaving $K$ invariant and that $X_+ \subeq X$ is a subset with the property that the 
kernel $K^\tau(x,y) := K(x, \tau  y)$ is also positive definite on $X_+$. 
We call such kernels $K$ {\it reflection positive} with respect to $(X,X_+, \tau)$. 
Then the closed subspace $\cE_+ \subeq \cE := \cH_K$ generated 
by $(K_x)_{x \in X_+}$ is $\theta$-positive for 
$(\theta f)(x) := f(\tau x)$. We thus obtain a 
reflection positive Hilbert space $(\cE,\cE_+,\theta)$. 

In this context, the space $\hat\cE$ can be identified with the reproducing kernel 
space $\cH^{K^\tau} \subeq \C^{X_+}$, where $q$ corresponds to the map 
\[ q \: \cE_+ \to \cH_{K^\tau}, \quad 
q(f)(x) := f(\tau(x))\] 
(\cite[Lemma~2.4.2]{NO18}). 
\end{example} 

For a symmetric semigroup $(G,S,\tau)$, 
we obtain natural classes of reflection positive kernels: 

\begin{defn}
A function $\vphi \colon G \to \C$ on a group $G$ is 
called {\it reflection positive} (\cite{JNO18}) if the kernel 
$K(x,y) := \vphi(xy^{-1})$ is reflection positive with respect to $(G,S,\tau)$ 
in the sense of Example~\ref{ex:1.3} 
with $X=G$ and $X_+=S$.
These are two simultaneous positivity conditions, namely that the kernel 
$\vphi(gh^{-1})_{g,h \in G}$ is positive definite on $G$ and that the kernel 
$\vphi(st^\sharp)_{s,t \in S}$ is positive definite on $S$.  
\end{defn}

The usal Gelfand--Naimark--Segal construction naturally extends to 
reflection positive functions and provides a correspondence with 
reflection positive representations (see \cite[Thm.~3.4.5]{NO18}). 

\begin{defn} \mlabel{def:1.4} 
For a symmetric semigroup  $(G,S,\tau)$, 
a unitary representation $U$ of $G$ on a reflection positive 
Hilbert space $(\cE,\cE_+,\theta)$ is called 
{\it reflection positive} if $\theta U_g \theta= U_{\tau(g)}$ for $g \in G$
 and $U_s\cE_+ \subeq \cE_+$ for every $s \in S$. 
\end{defn}

\begin{rem} \mlabel{rem:dil} 
(a)  If $(U_g)_{g \in G}$ is a reflection positive representation of 
$(G,S,\tau)$ on $(\cE,\cE_+,\theta)$, then we obtain contractions 
$(\hat U_s)_{s \in S}$ on $\hat\cE$, determined by 
\[  \hat U_s \circ q = q \circ U_s\res_{\cE_+} \quad \mbox { for } \quad 
s\in S,\] 
and this leads to an involutive representation $(\hat U,\hat\cE)$ of~$S$ by contractions  
(cf.\ \cite[Cor.~3.2]{JOl98}, \cite{NO14} or \cite{NO18}). 
We then call $(U,\cE,\cE_+,\theta)$ a {\it euclidean realization} 
of $(\hat U,\hat\cE)$.

(b) For $(G,S,\tau) = (\R,\R_+,-\id_\R)$, continuous reflection positive unitary one-parameter groups 
$(U_t)_{t \in \R}$ lead to a strongly continuous semigroup $(\hat U,\hat\cE)$ 
of hermitian contractions and every such semigroup $(C,\cH)$ has a natural 
euclidean realization obtained as the GNS representation associated to the positive definite 
operator-valued function $\vphi(t) := C_{|t|}$, $t \in \R$ (\cite[Prop.~6.1]{NO15a}).
\end{rem}

\begin{ex} \mlabel{ex:repo-realline} On $(\R,\R_+,-\id)$, we have: 
  \begin{itemize}
  \item[\rm(a)] For $0 \leq \alpha \leq 2$, the function $|x|^\alpha$ on $(\R,+)$ 
is  negative definite by \cite[Cor.~3.2.10]{BCR84} because 
$x^2$ is obviously negative definite. 
  \item[\rm(b)] For $\alpha \geq 0$, the function $|x|^\alpha$ is reflection negative if and only if 
   $0 \leq \alpha \leq 1$ (\cite[Ex.~4.3(a)]{JNO18}). 
  \item[\rm(c)] The function $-|x|^\alpha$ 
is reflection negative for $1 \leq \alpha \leq 2$ 
(\cite[Ex.~6.5.15]{BCR84}, \cite[Ex.~4.4(a)]{JNO18}). 
  \end{itemize}
\end{ex}

\section{Reflection positivity for affine actions} 
\mlabel{sec:4}

In this section we introduce reflection positive 
affine isometric actions $U \: G \to \Mot(\cE)$ on real Hilbert spaces~$\cE$ 
and relate it to the corresponding measure preserving action 
on the Gaussian $L^2$-space $\Gamma(\cE)$.

Let $(G,S,\tau)$ be a symmetric semigroup 
and $\cE$ be a real Hilbert space, 
endowed with an isometric involution $\theta$. 
We  consider an affine isometric action 
\begin{equation}
  \label{eq:aff-act}
\alpha_g v = U_g v + \beta_g \quad \mbox{ for } \quad g \in G, v \in \cE,
\end{equation}
where $U \colon G \to \OO(\cE)$ is an orthogonal representation 
and $\beta \colon G \to \cE$ a $1$-cocycle, i.e., 
\begin{equation}
  \label{eq:cocrel}
   \beta_{gh} = \beta_g + U_g \beta_h = \alpha_g \beta_h
 \quad \mbox{ for } \quad g,h \in G.
\end{equation}
Note that \eqref{eq:cocrel} in particular implies 
$\beta_e = 0$ and thus $\beta_{g^{-1}} = -U_g^{-1} \beta_g$. 
We further assume that $\theta \alpha_g \theta = \alpha_{\tau(g)},$ 
which is equivalent to 
\begin{equation}
  \label{eq:sym}
 \theta U_g \theta = U_{\tau(g)} \quad \mbox{ and } \quad 
\theta \beta_g = \beta_{\tau(g)} \quad \mbox{ for } \quad g \in G.
\end{equation}
If $\beta_G$ is total in $\cE$, then we can realize $\cE$ as a 
reproducing kernel Hilbert space $\cH_C \subeq \R^G$ with kernel 
\[ C(s,t) := \la \beta_s, \beta_t \ra, \qquad s,t \in G.\] 
For the function 
\[ \psi \: G \to \R, \quad \psi(g) := \|\beta_g\|^2 = C(g,g), \] 
we then obtain 
\begin{align*}
\psi(s^{-1}t) &= \|\beta_{s^{-1}t}\|^2 
= \|\beta_{s^{-1}} + U_s^{-1} \beta_t \|^2 
= \|U_s\beta_{s^{-1}} + \beta_t \|^2 
= \|\beta_t - \beta_s \|^2 
=\psi(s) + \psi(t) - 2 C(s,t),  
\end{align*}
so that 
\begin{equation}
  \label{eq:rel1}
 C(s,t) = \frac{1}{2}\big(\psi(s) + \psi(t) - \psi(s^{-1}t)\big) 
\quad \mbox{ and } \quad 
\psi(s^{-1}t) = C(s,s) + C(t,t) - 2 C(t,s). 
\end{equation}
In view of \eqref{eq:rel1}, 
$\psi$ is negative definite by Lemma~\ref{lem:bcr}. 
Equation \eqref{eq:rel1} implies that, if $\beta_G$ is total, then 
 the affine action $\alpha$ can be recovered completely from the function~$\psi$ 
and every real-valued negative definite function 
$\psi \: G \to \R$ with $\psi(e) = 0$ is of this form (cf.~\cite{Gu72, HV89}). 
We also note that $\theta\beta_g = \beta_{\tau(g)}$ implies 
that $\psi \circ \tau = \psi$. 

\begin{definition} \mlabel{def:7.1} (Reflection positive affine actions) 
The closed subspace $\cE_+$ generated by $(\beta_s)_{s \in S}$ 
is invariant under the affine action of $S$ on~$\cE$ 
because $\alpha_s \beta_t = \beta_{st}$ for $s,t \in S$. 
We call the affine action $(\alpha,\cE)$ {\it reflection positive} 
with respect to $(G,S,\tau)$ if $\cE_+$ is $\theta$-positive. 
\end{definition}

\begin{ex} (A universal example) \mlabel{ex:7.2}
Let $(\cE,\cE_+,\theta)$ be a reflection positive real Hilbert space, 
$\cE_- := \theta(\cE_+)$ and 
write $\Mot(\cE) \cong \cE \rtimes \OO(\cE)$ for 
its motion group. We  define an involution 
on $\Mot(\cE)$ by $\tau(b,g) := (\theta b, \theta g \theta)$. For 
$\gamma \in \Mot(\cE)$ we put $\gamma^\sharp :=\tau(\gamma)^{-1}$. Then 
\[ S 
:= \{ \gamma \in \Mot(\cE) \: \gamma\cE_+ \subeq \cE_+, \gamma^\sharp \cE_+ \subeq \cE_+ \} 
= \{ \gamma \in \Mot(\cE) \: \gamma\cE_+ \subeq \cE_+, \gamma \cE_- \supeq \cE_- \} \] 
is a $\sharp$-invariant subsemigroup of $\Mot(\cE)$ with 
\[ S \cap S^{-1} 
=\{ \gamma \in \Mot(\cE) \:  \gamma(\cE_+) = \cE_+, \gamma(\cE_-) = \cE_- \}.\] 
By construction, the affine action of 
$\Mot(\cE)$ on $\cE$ is reflection positive in the sense of 
Definition~\ref{def:7.1}.

For $\gamma = (b,g)$, the relation $\gamma(\cE_+) = \cE_+$ is equivalent to 
$b \in \cE_+$ and $g\cE_+ = \cE_+$. This shows that 
$(b,g) \in S \cap S^{-1}$ is equivalent to 
$b \in \cE_+ \cap \theta(\cE_+) = (\cE_+)^\theta$ (because of $\theta$-positivity) 
and to the condition that the restrictionss of $g$ to $\cE_\pm$ are unitary. 

The positive definite kernel $Q(x,y) := e^{-\|x-y\|^2/2}$ 
(Appendix~\ref{app:2}) is reflection positive with respect to 
$(G,S,\tau)$ because the kernel $Q^\theta(x,y) = Q(x, \theta y) = e^{-\|x-\theta y\|^2/2}$ 
is positive definite on $\cE_+$ (cf.~Example~\ref{ex:1.3}). 
From the $\Mot(\cE)$-invariance of $Q$, we thus obtain a reflection positive representation 
of $(\Mot(\cE),S,\tau)$ on the corresponding reflection positive Hilbert space 
$(\Gamma(\cE), \Gamma(\cE_+), \Gamma(\theta))$. 

It is instructive to make the corresponding space 
$\hat\Gamma(\cE)$ more explicit and to see how it identifies with 
$\Gamma(\hat\cE)$. 
From 
\[ \la \hat{e^{i\phi(v)}}, \hat{e^{i\phi(w)}} \ra
= \la e^{i\phi(v)}, \Gamma(\theta) e^{i\phi(w)} \ra
= \la e^{i\phi(v)}, e^{i\phi(\theta w)} \ra
= e^{-\frac{1}{2} \|v - \theta w \|^2}
= e^{-\frac{1}{2}(\|v\|^2 + \|w\|^2) + \la v, \theta w \ra} \] 
and 
\[ \la e^{i\phi(\hat v)}, e^{i\phi(\hat w)} \ra
= e^{-\frac{1}{2} \|\hat v - \hat w \|^2}
= e^{-\frac{1}{2}(\la v, \theta v\ra + \la w, \theta w \ra) + \la v, \theta w \ra} 
\quad \mbox{ in } \quad \Gamma(\hat\cE), \] 
we derive that 
\begin{equation}
  \label{eq:rp-fock}
e^{i\phi(\hat v)} = e^{\frac{1}{2}( \|v\|^2 - \la v, \theta v \ra)} \hat{e^{i\phi(v)}}
\quad \mbox{ and } \quad 
 \hat{e^{i\phi(v)}} = e^{\frac{1}{2}(\la v, \theta v \ra-  \|v\|^2)}e^{i\phi(\hat v)}.
\end{equation} 
For $\gamma \in S$, this leads to 
\[ \hat \gamma e^{i\phi(\hat v)}
= e^{\frac{1}{2}( \|v\|^2 - \la v, \theta v \ra)} \hat{e^{i\phi(\gamma v)}}
= e^{\frac{1}{2}( \|v\|^2 - \la v, \theta v \ra + \la \gamma v, \theta 
\gamma v\ra - \|\gamma v\|^2)}e^{i\phi(\hat{\gamma v})}.\] 
In particular, the cyclic subrepresentation generated by the 
constant function $1 = e^{i\phi(0)}$ is determined for $\gamma = (b,g)$ 
by the positive definite 
function 
\begin{align*}
\vphi(b,g) 
&= \la e^{i\phi(0)}, \hat{\gamma}e^{i\phi(0)} \ra 
= e^{\frac{1}{2}(\la b, \theta b\ra - \|b\|^2)} \la e^{i\phi(0)}, e^{i\phi(\hat{b})} \ra 
= e^{\frac{1}{2}(\la b, \theta b\ra - \|b\|^2)} e^{-\frac{1}{2} \|\hat b\|^2} \\
&= e^{\frac{1}{2}(\la b, \theta b\ra - \|b\|^2)} e^{-\frac{1}{2} \la b, \theta b \ra} 
= e^{-\frac{1}{2}\|b\|^2}.
\end{align*}
It follows that the function $\vphi(b,g) = e^{-\frac{1}{2}\|b\|^2}$ on $\Mot(\cE)$ is 
reflection positive for $(\Mot(\cE),S,\tau)$. 
\end{ex}

The following lemma provides a characterization of reflection positive 
affine actions in terms of kernels. 
\begin{lem} \mlabel{lem:repo} Let $(G,S,\tau)$ be a symmetric semigroup 
and $(\alpha, \cE)$ be an affine isometric action of $(G,\tau)$ 
on the real Hilbert space $\cE$. We write $\cE_+ := \oline{\Spann \beta_S}$ 
for the closed subspace generated by $\alpha_S(0)= \beta_S$. 
Then the following are equivalent:  
  \begin{itemize}
  \item[\rm(a)] The kernel $Q^\theta(x,y) = Q(x,\theta y) 
= e^{-\|x-\theta y\|^2/2}$ is positive definite on $\cE_+$. 
  \item[\rm(b)] $(\alpha,\cE)$ is reflection positive with respect to $(G,S,\tau)$, 
i.e., $\cE_+$ is $\theta$-positive. 
  \item[\rm(c)]  The kernel 
$C^\tau(s,t) :=  C(s,\tau(t))= \frac{1}{2}\big(\psi(s) + 
\psi(t) - \psi(s^\sharp t)\big)$ is positive definite on $(S,\sharp)$. 
  \item[\rm(d)]  The function $\psi\res_S \: S \to  \R$ is negative definite on $(S,\sharp)$. 
  \end{itemize}
\end{lem}

\begin{prf} (a) $\Leftrightarrow$ (b): In view of 
$Q(x, \theta y) = e^{-\frac{\|x\|^2}{2}} e^{-\frac{\|\theta  y\|^2}{2}} e^{\la x, \theta  y \ra}
= e^{-\frac{\|x\|^2}{2}} e^{-\frac{\|y\|^2}{2}} e^{\la x, \theta  y \ra},$ 
the kernel $Q^\theta$ is positive definite on $\cE_+$ if and only if 
the kernel $e^{\la x, \theta y \ra}$ is positive definite on $\cE_+$, but this is 
equivalent to $\cE_+$ being $\theta$-positive (\cite[Rem.~2.8]{NO15a}). 

(b) $\Leftrightarrow$ (c): Since $\cE_+$ is generated by 
$(\beta_s)_{s \in S}$, this follows from \eqref{eq:rel1} and the definition 
of~$C$. 

(c) $\Leftrightarrow$ (d): By Lemma~\ref{lem:bcr},  
the kernel $C^\tau$ is positive definite if and only if the kernel 
$(\psi(s^\sharp t))_{s,t \in S}$ is negative definite, which is (d). 
\end{prf}
 
This leads us to the following concept: 

\begin{definition} \mlabel{def:repo-aff}
We call a continuous function $\psi \colon G \to \R$ {\it reflection negative} 
with respect to $(G,S,\tau)$ if 
$\psi$ is a negative definite function on $G$ 
and $\psi\res_S$ is a negative definite function on the involutive semigroup 
$(S,\sharp)$ (Definition~\ref{def:8.1.1}). 
\end{definition}

From Schoenberg's Theorem for kernels (Remark~\ref{rem:schoen}) 
we immediately obtain from Lemma~\ref{lem:repo}: 

\begin{cor} Let $(\alpha,\cE)$ be a reflection positive affine action 
of $(G,S,\tau)$. Then, for every $h > 0$, the function 
$\vphi_h(g) := e^{-h \|\beta_g\|^2}$ is reflection positive, i.e., 
the function $\|\beta_g\|^2$ is reflection negative. 
\end{cor}

\begin{rem} (a) Let $\cH$ be a real Hilbert space. 
For $h > 0$, the function $\vphi_h(b,g) := e^{-h\|b\|^2}$ on $\Mot(\cH)$ 
is positive definite. A corresponding cyclic representation can be realized 
as follows. We consider the unitary representation of $\Mot(\cH)$ 
on $L^2(\cH^*,\gamma_h)$  given by 
\[ \rho(b,g)F= e^{i \phi(b)} g_* F,
\quad \mbox{ i.e.} \quad 
(\rho(b,g)F)(\alpha) = e^{i \alpha(b)} F(\alpha \circ g), \quad \alpha \in \cH^*,
\] 
where $\gamma_h$ is the Gaussian measure on $\cH^*$ with Fourier transform 
$\hat\gamma_h(v) = e^{-h\|v\|^2}$ 
and $\phi(v)(\alpha) = \alpha(v)$ as in 
Definition~\ref{def:3.5} (see also Remark~\ref{rem:unirep-mot}). 
Then the constant function $1$ is a cyclic vector, and the corresponding positive definite 
function is 
\begin{equation}
  \label{eq:posdef-fun}
\la 1, \rho(b,g)1 \ra = \E(e^{i\phi(b)}) = e^{-h\|b\|^2} = \vphi_h(b,g).
\end{equation}

(b) We conclude that, for every reflection positive affine action $(\alpha,\cE)$,  
for $(G,S,\tau)$, a cyclic reflection positive representation 
of $(G,S,\tau)$ corresponding to 
$\vphi_h(g) = e^{-h\|\beta_g\|^2}$ is obtained on the cyclic subspace of 
$L^2(\cE^*,\gamma_h)$ generated by the constant function~$1$.
\end{rem}

\section{Some unitary representations of $\GL_2(\R)$} 
\mlabel{sec:5}

In this section we introduce a family of unitary 
representations $(U^H, \cH_H)_{0 < H < 1}$ of $\GL_2(\R)$, 
respectively of the projective group $\PGL_2(\R)\cong \GL_2(\R)/\R^\times$. \\

We identify the real projective line 
$\bP_1(\R) \cong \bS^1$ of one-dimensional linear subspaces of $\R^2$ 
with $\R_\infty = \R \cup \{\infty\}$. On this space the group 
$G := \GL_2(\R)$ acts naturally by fractional linear maps  
\[ g.x = g(x) = \frac{a x + b}{cx + d} \quad \mbox{ for } \quad 
g = \pmat{a & b \\ c & d}.\] 
Note that 
\begin{equation}
  \label{eq:gprime}
g'(x) = \frac{ad-bc}{(cx + d)^2},
\end{equation}
which shows that $g$ acts on the circle $\R_\infty$ 
in an orientation preserving fashion if and only if 
$\det g > 0$.  

\begin{defn} For two different elements $\alpha \not= \gamma \in \R_\infty$, 
we write $(\alpha,\gamma)$ for the open interval between 
$\alpha$ and $\gamma$ with respect to the cyclic order. For 
$\gamma < \alpha$ in $\R$ this means that 
\[ (\alpha,\gamma) = (\alpha, \infty) \cup \{\infty\} \cup (-\infty, \gamma).\] 
\end{defn}

\begin{defn} For the action of $G$ on $\R_\infty$, 
Lebesgue measure $\lambda$ on $\R$, resp., the corresponding 
measure on $\bS^1 \cong \R_\infty$ with $\lambda(\{\infty\}) = 0$  
is quasi-invariant with 
$\frac{d (g^{-1}_*\lambda)}{d\lambda}(x) = |g'(x)|.$ 
A  unitary representation of
$\GL_2(\R)$ (resp., of $\PGL_2(\R)$)  on $L^2(\R) = L^2(\R,\lambda)$ is given by 
\begin{equation}
  \label{eq:unirep}
(U_g\xi)(x) = \sgn(\det g)
\frac{|ad-bc|^{1/2}}{|cx + d|} \xi\Big(\frac{ax + b}{cx + d}\Big) \quad \mbox{ for } \quad g^{-1} = \pmat{a & b \\ c &d}.
\end{equation} 
We could as well work without the $\sgn(\det g)$-factor, 
but we shall see below that it is more natural this way when it comes to 
the relation with fractional Brownian motion. 
\end{defn} 

We now explain how this representation can be embedded into a 
family of unitary representations $(U^H)_{0 < H < 1}$. 
For $H \not= \shalf$, these representations belong to the so-called 
{\it complementary series} (cf.~\cite{vD09}, \cite{JO00}, \cite{NO14}). 
For $H > \shalf$, the corresponding 
Hilbert space $\cH_H$ is the completion of the Schwartz space $\cS(\R)$ 
with respect to the inner product 
\begin{equation}
  \label{eq:scalpro-H}
\la \xi, \eta \ra_H := H(2H-1) \int_{\R} \int_{\R} 
\oline{\xi(x)}\eta(y) \frac{dx\ dy}{|x-y|^{2-2H}}.
\end{equation}
Note that $2 - 2H \in (0,1)$, so that the kernel $|x-y|^{2H-2}$ is locally 
integrable and defines a positive definite distribution kernel on~$\R$. 
This implies in particular that \eqref{eq:scalpro-H} 
makes sense for any pair of compactly supported bounded measurable functions 
on $\R$ and that any such function defines an element of $\cH_H$. 
In  Appendix~\ref{app:d} we show that
\begin{equation}
  \label{eq:newscal-a}
 \la \xi,\eta\ra_H 
= -\frac{1}{2} \int_\R \int_\R \oline{\xi'(x)}\eta'(y) |x-y|^{2H}\, dx\, dy. 
\end{equation}

\begin{defn}
As we have seen in Example~\ref{ex:repo-realline}(a), 
the continuous function $D^H(x) = |x|^{2H}$ on $\R$ is negative definite for 
$0 < H \leq 1$. 
Therefore \eqref{eq:newscal-a} defines for $0 < H < 1$ a positive semidefinite 
form on $\cS(\R)$. We write $\cH_H$ for the corresponding Hilbert space. 
Here we use that the total integrals of $\xi'$ and $\eta'$ vanish 
(cf.\ Remark~\ref{rem:newscal-schwartz}). 
Note that this definition also makes sense for $H = 0$ and $H =1$, 
but $\cH_0 = \{0\}$ and $\cH_1$ is one-dimensional. 
 \end{defn}

\begin{defn}
We obtain unitary representations of
$\GL_2(\R)$ (resp., the quotient $\PGL_2(\R)$) on $\cH_H$, $0 < H < 1$ by 
\begin{equation}
  \label{eq:unirepH}
(U_g^H\xi)(x) = \sgn(\det g)\frac{|ad-bc|^{H}}{|cx + d|^{2H}} 
\xi\Big(\frac{ax + b}{cx + d}\Big) \quad \mbox{ for } \quad g^{-1} = \pmat{a & b \\ c &d}.
\end{equation}
For the verification of unitarity we refer to Appendix~\ref{app:d}. 
For $H = \shalf$, we obtain the representation 
on $L^2(\R) \cong \cH_{1/2}$ from \eqref{eq:unirep}. 
\end{defn}

\begin{rem} (a) Considering the singularities of the factors in 
the formula for $U^H_g$, we see that the operators 
$U^H_g$ preserve the class of locally bounded measurable functions for 
which 
\[  \sup_{x \in \R} |x|^{2H} |\xi(x)| < \infty.\] 
For $H > \shalf$, all these functions are contained in $\cH_H$, 
so that we obtain a dense subspace of $\cH_H$ invariant under the operators~$U^H_g$.   

(b) We note that the representation $(U^H, \cH_H)$ 
is equivalent to $(U^{1-H}, \cH_{1-H})$, as can be seen 
by realizing these representations on $\bS^1$ (see \cite[Ch.~7]{NO18}). 
We will not use this duality here. 
\end{rem}

\begin{rem}
The unitary representations $(U^H)_{0 <  H < 1}$ 
of $\GL_2(\R)$ yield in particular three important one-parameter groups:
\begin{itemize}
\item Translations: $(S_t^H \xi)(x) = \xi(x-t)$ for $t \in \R$. 
\item Dilations: $(\tau_a^H \xi)(x) = \sgn(a) |a|^H \xi(ax)$ for 
$a \in \R^\times$. 
\item Inverted translations: 
$(\kappa_t^H \xi)(x) = \frac{1}{|1 - tx|^{2H}} \xi(\frac{x}{1- tx})$ 
 for $t \in \R$. 
\end{itemize}
Note that 
\begin{equation}
  \label{eq:trans-dil}
\tau_{r^{-1}}^H S_t^H \tau_r^H = S^H_{rt} \quad \mbox{ for } \quad t \in\R, 
r \in \R^\times.
\end{equation}
For $\sigma = \pmat{0 & 1 \\ 1 & 0} \in \GL_2(\R)$ 
with $\sigma.x = \frac{1}{x}$, 
we have 
\begin{equation}
  \label{eq:sigma-act}
U_\sigma^H(\xi)(x)  = -|x|^{-2H} \xi(x^{-1}) 
\quad \mbox{ and } \quad 
 U^H_\sigma S^H_t U^H_\sigma =  \kappa_t^H.
\end{equation}
\end{rem}

\section{Fractional Brownian motion} 
\mlabel{sec:6}

In this section we introduce fractional Brownian motion in 
terms of its covariance kernel. We then show that the unitary 
representations $(U^H)_{0 < H < 1}$ of $\GL_2(\R)$ and a realization 
of fractional Brownian motion in the Hilbert space $\cH_H$,
 resp., on its Fock space, can be used to obtain 
in a very direct and simple fashion the projective invariance of 
fractional Brownian motion. 

\subsection{A realization of fractional Brownian motion} 

\begin{defn} \mlabel{def:4.1}
{\it Fractional Brownian motion with Hurst index $H \in (0,1)$}  
is a real-valued Gaussian process 
$(B^H_t)_{t \in \R}$ with zero means and covariance kernel 
\[ C^H(s,t) = \E(B_s^H B_t^H) = \frac{1}{2}(|s|^{2H} + |t|^{2H} - |s-t|^{2H}) 
\quad \mbox{ for } \quad s,t \in \R\] 
(cf.~\cite[Satz~7]{Ko40a} for the determination of those parameters 
for which this kernel is positive definite). A curve 
$\gamma \: \R \to \cH$ with values in a Hilbert space $\cH$ satisfying 
$\la \gamma(s),\gamma(t) \ra = C^H(s,t)$ is called a {\it fractional 
Wiener spiral}. 

Brownian motion arises for $H = 1/2$, 
and in this case 
\[ C^{1/2}(s,t) = \frac{1}{2}(|s| + |t| - |s-t|) 
=\begin{cases}
|t| \wedge |s| & \text{ for } st \geq 0. \\
0 & \text{ for } st < 0.  
\end{cases}
\]
\end{defn}

We refer to the monograph \cite{BOZ08} for a 
stochastic calculus for fractional Brownian motion.

\begin{ex} \mlabel{ex:bifrac} 
(Bifractional Brownian motion) 
For $0 < H \leq 1$ and $0 < K \leq 1$, the kernel 
\[ C(s,t) = (|t|^{2H} + |s|^{2H})^K - |t-s|^{2HK} \] 
on $\R$ is positive definite (Lemma~\ref{lem:bcr}). 
The corresponding centered Gaussian 
process $B^{H,K}$ is called {\it bifractional Brownian motion} 
(\cite{HV03}). 
For $K = 1$ we obtain fractional Brownian motion which has 
stationary increments, but for $K < 1$ the process $B^{H,K}$ 
does not have this property 
since the kernel 
\[ D(t,s) = C(t,t) + C(s,s) - 2 C(t,s) 
= 2^K (|t|^{2HK} + |s|^{2HK}) - (|t|^{2H} + |s|^{2H})^K + |t-s|^{2HK} \] 
on $\R$ is not translation invariant. 

For a concept of {\it trifractional Brownian motion} and 
decompositions of fractional Brownian motion into independent 
bifractional and trifractional components we refer to \cite{Ma13}. 
\end{ex}

\begin{rem} \mlabel{rem:ch-trafo}
For $0 < H < 1$, the kernel $C^H$ satisfies 
\begin{equation}
  \label{eq:ctrafo}
C^H(\lambda s, \lambda t) = |\lambda|^{2H} C(s,t) 
\quad \mbox{ and }\quad 
C^H(s^{-1}, t^{-1}) = |st|^{-2H} C^H(s,t) 
\quad \mbox{ for }\quad s,t \in \R^\times, \lambda \in \R.
\end{equation}

These transformation rules show that: \\

(a)  For a fractional Brownian motion $(B^H_t)_{t \in \R}$ with Hurst index~$H$, 
the centered Gaussian process $(X_t)_{t \in \R}$ defined by 
\[ X_0 := 0 \quad \mbox{ and }\quad 
 X_t := |t|^{2H} B_{1/t} \quad \mbox{ for } \quad t \not= 0\] 
also is a fractional Brownian motion with Hurst index~$H$. 

(b) For $c \in \R^\times$ and 
$X_t := |c|^{-H} B_{ct}^H,$ the process $(X_t)_{t \in \R}$ is a 
fractional Brownian motion with Hurst index~$H$. 

(c) For $h \in \R$, the process 
$X_t := B^H_{t+h} - B^H_h$ also is a 
fractional Brownian motion with Hurst index~$H$. 
\end{rem}

\begin{lem} For $t \in \R$ and $0 < H < 1$, consider the random 
variables 
\[ B_t^H = \phi(b_t^H) \quad \mbox{ for }\quad  
b_t^H := \sgn(t) \chi_{[t \wedge 0, t \vee 0]} 
= \chi_{[0,\infty)} - \chi_{[t,\infty)} \in \cH_H.\] 
Then $\la b_s^H, b_t^H \ra_{H} = C^H(s,t)$, i.e., 
$(B_t^H)_{t \in \R}$ is a realization of fractional 
Brownian motion with Hurst index~$H$.
\end{lem}

\begin{prf} {\bf Case $H \geq \shalf$:} 
As $C^H(s,t) = C^H(-s,-t) = C^H(t,s)$, we only have to show that 
\[ C^H(s,t) = H (2H-1) \int_0^t \int_0^s \frac{dx\, dy}{|x-y|^{2 - 2H}}
\quad \mbox{ for } \quad 0 < s \leq t \] 
and  
\[ C^H(s,t) = -H (2H-1) \int_t^0 \int_0^s \frac{dx\, dy}{|x-y|^{2 - 2H}}
\quad \mbox{ for } \quad t < 0 < s.\] 
This is an elementary calculation. 

{\bf Case $H < \shalf$:} In this case we can calculate the 
scalar product \eqref{eq:newscal} by using the formula 
$(b^H_t)' = \delta_0 - \delta_t$ (a difference of two point evaluations). 
This leads to 
\begin{align*}
\la b^H_s, b^H_t \ra_H 
&= - \frac{1}{2} \int_\R (b^H_s)'(y)(|y|^{2H} - |t - y|^{2H})\, dy \\
&= - \frac{1}{2}\big(-|t|^{2H} -(|s|^{2H} - |t-s|^{2H})) 
=  \frac{1}{2}\big(|t|^{2H} + |s|^{2H} - |t-s|^{2H}).  
\qedhere\end{align*}
\end{prf}

For $H > \shalf$, the preceding lemma follows from 
\cite[p.~168]{AL08} and for $H = \shalf$ it is already 
contained in \cite[p.~117]{Ko40a}. 
Other realizations of fractional Brownian motion are discussed in \cite{AJP12}. 

\begin{rem} For Brownian motion $(H = \shalf)$, 
an alternative realization is obtained by 
$b_t' := \chi_{[t \wedge 0, t \vee 0]}$ for $t \in \R$ (see \cite[p.~130]{Hid80}).  
For $H\not= 1/2$ this sign change does no longer work because 
$C^H(s,t) \not=0$ for $st < 0$. 
\end{rem}

\subsection{Projective invariance of the covariance kernels} 

Recall the {\it cross ratio} 
\[ \CR(z, z_1, z_2, z_3) := \frac{(z-z_1)(z_2 - z_3)}{(z-z_3)(z_2 - z_1)} 
\quad \mbox{ of four different elements } \quad z,z_1,z_2, z_3 \in \R_\infty \] and that it is invariant 
under the action of $\GL_2(\R)$. As 
$\CR(z, 0,1,\infty) = z$, we obtain 
for $g(\alpha, \beta, \gamma) = (0,1,\infty)$ the relation 
\begin{equation}
  \label{eq:g-cr}
g(z) = \CR(g(z), 0,1,\infty) = \CR(z,\alpha, \beta, \gamma) 
= \frac{z-\alpha}{z-\gamma} \cdot \frac{\beta-\gamma}{\beta-\alpha},   
\end{equation}
expressing $g$ as a cross ratio. Accordingly, we obtain 
for each triple $(\alpha,\beta, \gamma)$ of mutually different elements 
of $\R_\infty$ the following kernel 
\begin{eqnarray}
  \label{eq:abc-kernel}
C^H_{\alpha,\beta,\gamma}(s,t) 
&:=& C^H(g(s),g(t)) 
= C^H\Big(
\frac{(s-\alpha)(\beta-\gamma)}{(s-\gamma)(\beta-\alpha)}, 
\frac{(t-\alpha)(\beta-\gamma)}{(t-\gamma)(\beta-\alpha)}\Big) \\
&=& \Big|\frac{\beta-\gamma}{\beta-\alpha}\Big|^{2H} 
C^H\Big(\frac{s-\alpha}{s-\gamma},\frac{t-\alpha}{t-\gamma}\Big),\notag
\end{eqnarray}
where the last expression only makes sense for 
$\alpha,\beta, \gamma\in\R$. 
By construction we then have 
\begin{equation}
  \label{eq:transfo}
C^H_{h.\alpha,h.\beta,h.\gamma}(h(s),h(t)) = C^H_{\alpha,\beta,\gamma}(s,t) 
\quad \mbox{ for }\quad h \in \GL_2(\R), 
s,t \not= \gamma.
\end{equation}
Note that $C^H_{0,1,\infty} = C^H$ and that, 
for $\beta\in \R^\times$, $\alpha = 0$ and $\gamma = \infty$, we obtain in particular 
for the dilation $g.x = \beta^{-1}x$: 
\[ C^H_{0,\beta,\infty}(s,t) 
= C^H(\beta^{-1}s,\beta^{-1} t) 
= |\beta|^{-2H} C^H(s,t),\]  
which is a multiple of $C^H$. In particular, 
normalization of $C^H$ and $C^H_{0,\beta,\infty}$ leads on 
$\R^\times$ to the same kernels. We also observe that 
\[ C^H_{\infty,\beta,0}(s,t) 
= C^H\Big(\frac{\beta}{s}, \frac{\beta}{t}\Big) 
=\frac{|\beta|^{2H}}{|st|^{2H}}  C^H(s, t) \] 
implies the equality of the normalized kernels 
$\tilde C^H_{\infty,\beta,0}(s,t) 
=  \tilde C^H_{0,\beta,\infty}(s,t).$ 

From \eqref{eq:abc-kernel} and the preceding discussion 
we obtain immediately: 
\begin{lem} \mlabel{lem:5.6} 
For $\alpha\not=\gamma$ in $\R_\infty$, the normalized kernel 
$C^H_{\alpha,\gamma} := \tilde C^H_{\alpha,\beta,\gamma}$ on 
$\R_\infty \setminus \{\alpha,\gamma\}$ 
does not depend on~$\beta$ and satisfies the symmetry condition  
$C^H_{\alpha,\gamma} = C^H_{\gamma,\alpha}.$ 
\end{lem}

\begin{prop} \mlabel{prop:kern-invar} 
For $g \in \GL_2(\R)$ we have 
\begin{equation}
  \label{eq:5.7}
C^H_{g.\alpha, g.\gamma}(g(s),g(t)) = C^H_{\alpha,\gamma}(s,t)
\quad \mbox{ for } \quad s,t \not\in \{\alpha,\gamma\}.
\end{equation}
In particular, if $g \in \GL_2(\R)$ preserves the $2$-element set $\{\alpha,\gamma\}$, then 
the kernel $C^H_{\alpha,\gamma}$ on $\R_\infty \setminus \{\alpha,\gamma\}$ 
is $g$-invariant. 
\end{prop}

\begin{prf} Equation \eqref{eq:5.7} follows directly from 
\eqref{eq:transfo} and the remainder is a consequence 
of Lemma~\ref{lem:5.6}.  
\end{prf}

The preceding proposition expresses the 
projective invariance of fractional Brownian motion in the sense of 
P.~L\'evy. For $H = 1/2$, this is classical 
(\cite[\S I.2]{Le65}, \cite{HKNY69}, \cite[Thm.~5.2]{Hid80}). 

\begin{rem} The identity component $G^{\alpha,\gamma}_0$ of the 
stabilizer $G^{\alpha,\gamma}$ of $\{\alpha,\gamma\}$ in $G = \PGL_2(\R)$ 
is a (hyperbolic) one-parameter group of 
$\SL_2(\R)$ whose fixed points are $\alpha$ and $\beta$ 
(these are the orientation preserving transformations mapping 
the interval $(\alpha,\gamma)$ onto itself). 
The full stabilizer of the pair $(\alpha,\gamma)$ 
is isomorphic to $\R^\times$. It also contains 
an involution in $\PSL_2(\R)$ exchanging the two connected components of 
$\R_\infty \setminus \{\alpha,\gamma\}$. 

Moreover, there exists for 
each $\beta \not=\alpha,\gamma$ a unique involution 
$\theta^{\alpha,\beta,\gamma} \in \GL_2(\R)$ exchanging 
$\alpha$ and $\gamma$ and fixing $\beta$. 
It satisfies 
\[  \theta^{\alpha,\beta,\gamma} g \theta^{\alpha,\beta, \gamma} = g^{-1} 
\quad \mbox{ and }\quad 
g\theta^{\alpha,\beta,\gamma} g^{-1} =  \theta^{\alpha,g.\beta, \gamma} 
\quad \mbox{ for } \quad 
g \in G^{\alpha,\gamma}_0.\] 
The subgroup $G^{\alpha,\gamma} \subeq \PGL_2(\R)$ has four connected components. 
\end{rem}

\subsection{Projective invariance of the realization} 

We now link the projective invariance of fractional Brownian motion to the 
specific realization in the Hilbert space $\cH_H$. 
Formula (b) in the theorem below connects the normalized 
projective transforms of the kernel $C^H$ to the unitary 
representation $U^H$ of $\GL_2(\R)$ on $\cH_H$. 

\begin{thm} \mlabel{thm:f-ker-rel}
For a triple $(\alpha,t,\gamma)$ of mutually different 
points in $\R_\infty$, there exists a uniquely determined 
M\"obius transformation $g_t \in \PSL_2(\R)$ with 
$(g_t(\alpha), g_t(t),g_t(\gamma))  = (0,1,\infty).$
We thus obtain functions of the form 
\begin{equation}
  \label{eq:3-point-random-fcts}
 f^{\alpha,\gamma}_t(x) := \big(U_{g_t^{-1}}^H \chi_{[0,1]}\big)(x) 
= \sgn(\det g_t)\frac{|\det g_t|^H }{|cx + d|^{2H}} \chi_{g_t^{-1}([0,1])}(x) 
\quad \mbox{ for } \quad t \in \R_\infty \setminus \{\alpha,\gamma\}.
\end{equation}
Then the following assertions hold: 
\begin{itemize}
\item[\rm(a)] All functions $f^{\alpha,\gamma}_t$ are unit vectors in $\cH_H$.
\item[\rm(b)] 
$\la f_s^{\alpha,\gamma}, f_t^{\alpha,\gamma} \ra_{\cH_H} 
= C^H_{\alpha,\gamma}(s,t)$ for $s,t \not\in \{\alpha,\gamma\}.$
\end{itemize}
\end{thm}

\begin{prf} (a)  
As $\|\chi_{[0,1]}\|_{\cH_H} = 1$ and the representation $U^H$ is unitary, 
the functions $f^{\alpha,\gamma}_t$ are unit vectors in $\cH_H$. 

(b) For $g = \pmat{ t & 0 \\ 0 & 1}$ with $g.x = tx$, 
we have  
\begin{equation}
  \label{eq:dilat}
U_g \chi_{[0,1]} =  |t|^{-H}  b^H_t, \quad \mbox{ resp.} \quad 
b^H_t = |t|^H  \cdot U_g \chi_{[0,1]} 
\quad \mbox{ for } \quad 
t \in \R^\times.
\end{equation}
This relation is the reason for the $\sgn(\det g)$-factor in the definition of~$U^H$.

For $s,t\not\in \{ \alpha,\gamma\}$, the element 
$g_s g_t^{-1}$ fixes $0$ and $\infty$, hence is linear and given by 
multiplication with $(g_s g_t^{-1})(1)= g_s(t)$. 
We thus obtain with Remark~\ref{rem:ch-trafo}, \eqref{eq:g-cr} and \eqref{eq:dilat}  
\begin{align*}
\la f_s^{\alpha,\gamma}, f_t^{\alpha,\gamma} \ra_{\cH_H} 
&= \la U^H_{g_s^{-1}} \chi_{[0,1]}, U^H_{g_t^{-1}} \chi_{[0,1]} \ra_{\cH_H} 
= \la \chi_{[0,1]}, U^H_{g_s g_t^{-1}} \chi_{[0,1]} \ra_{\cH_H} \\
&= \la b^H_1, |g_s(t)|^{-H} b^H_{g_s(t)}\ra_{\cH_H} 
= |g_s(t)|^{-H} C^H(1, g_s(t))\\ 
&= \frac{|t - \gamma|^H |s-\alpha|^H} {|t - \alpha|^H |s-\gamma|^H} 
C^H\Big(1, \frac{(t-\alpha)(s-\gamma)}{(t- \gamma)(s-\alpha)}\Big) \\ 
&= \frac{|t - \gamma|^H |s-\gamma|^H} {|t - \alpha|^H |s-\alpha|^H} 
C^H\Big(\frac{s-\alpha}{s- \gamma}, \frac{t- \alpha}{t - \gamma}\Big)\\
&= \frac{|t - \gamma|^H |s-\gamma|^H} {|t - \alpha|^H |s-\alpha|^H} 
\frac{|\beta-\alpha|^{2H}}{|\beta-\gamma|^{2H}} 
C^H_{\alpha,\beta,\gamma}(s,t).
\end{align*}
Since the kernel on the left hand side is normalized, (b) follows. 
\end{prf}

From Proposition~\ref{prop:kern-invar} and 
Theorem~\ref{thm:f-ker-rel}, we obtain: 

\begin{cor} \mlabel{cor:proj-inv}  
The normalized stochastic process defined by $(f^{\alpha,\gamma}_t)_{t\not=\alpha, \gamma}$ 
is stationary with respect to 
the stabilizer of the two-point set $\{\alpha,\gamma\}$ in $\GL_2(\R)$.   
\end{cor}

\begin{rem} \mlabel{rem:5.x} 
Since the representations $(U^H, \cH_H)$ of $\GL_2(\R)$ are irreducible, 
\cite[Prop.~5.20]{JNO16a} implies that the space 
$\cH_H^\infty$ of smooth vectors is nuclear. 
Therefore \cite[Cor.~5.19]{JNO16a} shows that the Gaussian measure 
$\gamma_{\cH_H}$ can be realized on the space $\cH_H^{-\infty}$ of distribution vectors 
for this representation (cf.\ Appendix~\ref{app:2}). 
Therefore our construction leads to a realization
 of fractional Brownian motion on the topological dual space $\cH_H^{-\infty}$ 
of the $\GL_2(\R)$-invariant subspace $\cH_H^\infty$ of smooth vectors. 

From the proof of \cite[Prop.~5.20(b)]{JNO16a}, we further derive that 
an element $\xi \in \cH_H$ is a smooth vector if and only if it is a smooth 
vector for the compact subgroup $K = \OO_2(\R)$. Considering 
$\cH_H$ as a space of distributions on the circle $\bS^1$, it is not hard to see 
that $\cH_H^\infty = C^\infty(\bS^1)$ and hence that 
$\cH_H^{-\infty} = C^{-\infty}(\bS^1)$ is the space of distributions on the circle. 
  
\end{rem}

\section{Fractional Brownian motion and reflection positivity} 
\mlabel{sec:7}

We now turn to reflection positivity in connection 
with fractional Brownian motion. 
Our main result is Theorem~\ref{thm:6.3} 
on the reflection positivity of the normalized kernels $C^H_{\alpha,\gamma}$ 
for $H \leq \shalf$. 
We start with the normalization of the kernel 
$C^H$, which corresponds to the pair $(\alpha,\gamma) = (0,\infty)$.

\begin{prop} \mlabel{prop:6.1} The kernel 
\[ C^H_{0,\infty}(s,t) = \tilde C^H(s,t) = \frac{C^H(s,t)}{|s|^H |t|^H}
\quad \mbox{ on } \quad X := \R^\times \] 
is invariant under the involution $\theta(x) = x^{-1}$. 
It is reflection positive on $X_+ := (-1,1) \cap \R^\times$ if and only 
of $0 < H \leq \shalf$. If this is the case, then 
$\hat\cE \cong L^2((0,\infty),\mu)$, with the measure 
\[ \mu = \delta_{2H} + \sum_{k = 1}^\infty {2H \choose k} (-1)^{k-1} \delta_k.\] 
For $H = \shalf,$ we have $\mu = 2\delta_1$, and $\hat\cE$ is one-dimensional. 
\end{prop}

\begin{prf}  Reflection positivity with respect to 
$(X,X_+,\theta)$ is equivalent to the positive definiteness of the 
kernel 
\[ C^H_{0,\infty}(s, t^{-1}) 
= |t|^H |s|^{-H} C^H(s,t^{-1})
= |t|^{-H} |s|^{-H} C^H(st,1) 
= \frac{1}{2|t|^H |s|^H}(1 + |st|^{2H} - (1 - st)^{2H}) \] 
for $|t|, |s| < 1$ (cf.\ Example~\ref{ex:1.3}).
This kernel is positive definite on $(-1,1)$ if and only if the function 
\[ \vphi(t) := 1 + t^{2H} - (1 - t)^{2H} 
= t^{2H} - \sum_{k = 1}^\infty {2H \choose k} (-1)^k t^k \] 
on the multiplicative semigroup $S = ((0,1),\id)$ is 
positive definite. For $k \geq 1$ we have 
\[ (-1)^{k-1} {2H \choose k} 
= \frac{2H(1-2H)(2-2H)\cdots (k-1-2H)}{k!}.\] 
For $\shalf < H < 1$, we have $1 - 2H < 0$ and all other factors are positive. 
As $\vphi$ is increasing, it is positive definite if and only if there exists a 
positive Radon measure $\mu$ on $[0,\infty)$ with 
\[ \vphi(t) = \int_0^\infty t^\lambda\, d\mu(\lambda)\] 
(use \cite[Prop.~4.4.2]{BCR84} or apply 
\cite[Cor.~VI.2.11]{Ne00} to $S \cong ((0,\infty),+)$). 
Therefore $\vphi$ is positive definite if and only if $H \leq \shalf$. 
In this case the description of $\hat\cE$ 
follows from the proof of \cite[Thm.~VI.2.10]{Ne00}.  
\end{prf}

\begin{rem} (Reflection positivity of fractional Brownian motion) 
For $\theta = \pmat{0 & 1 \\ 1 & 0}$ with $\theta.x = \theta^{0,1,\infty}(x) 
= x^{-1}$,  we have $(U^{1/2}_{\theta}\xi)(x) = -|x|^{-2H} \xi(x^{-1}).$ 
In particular, 
\[  (U^H_{\theta} \chi_{[0,t]})(x) = -|x|^{-2H}  \chi_{[t^{-1},\infty)}.\] 
Therefore $U^H_{\theta}$ is not the unique unitary involution 
$\hat\theta$ of $\cH_H$ transforming $b_t^H$ into $|t|^{2H} b_{1/t}$ 
(cf.\ Remark~\ref{rem:ch-trafo}).
\end{rem}

With the kernels $C^H_{\alpha,\gamma}$ (Lemma~\ref{lem:5.6}), we 
obtain a family of normalized Gaussian 
processes, covariant with respect to the action of 
$\GL_2(\R)$ on $\R_\infty$. The following proposition shows that, 
for $H \leq \shalf$, these kernels are reflection 
positive with respect to involutions exchanging $\alpha$ and $\gamma$. 

\begin{thm} \mlabel{thm:6.3} 
Let $\alpha, \beta, \gamma$ in $\R_\infty$ be mutually different 
and let $\theta := \theta^{\alpha,\beta,\gamma}$ 
be the projective involution exchanging $\alpha$ and $\gamma$ and fixing 
$\beta$. Let $X := \R_\infty \setminus \{\alpha,\gamma\}$ and 
$X_\pm \subeq X$ be the intersection of $X$ with the two connected components 
of the complement of the fixed point set of $\theta$ (which consists of two points). 
Then the kernel $C^H_{\alpha,\gamma}$ 
is reflection positive with respect to~$(X,X_+,\theta)$ if and only if 
$H \leq \shalf$. 
\end{thm}

\begin{prf} Since the family of kernels $C^H_{\alpha,\gamma}$ is invariant under the 
action of $\GL_2(\R)$ and 
\[ g \theta^{\alpha,\beta,\gamma} g^{-1}
= \theta^{g.\alpha, g.\beta,g.\gamma},\] 
it suffices to verify the assertion 
for $(\alpha,\beta, \gamma) = (0,1,\infty)$. Then $\theta(x) = x^{-1}$ and 
we may put $X_+ = (-1,1)\setminus \{0\}$. 
Hence the assertion follows from  Proposition~\ref{prop:6.1}. 
\end{prf}

\begin{ex} For $(\alpha,\beta,\gamma) = (-1,0,1)$, the involution 
$\theta := \theta^{\alpha,\beta,\gamma}$ is given by $\theta(x) = -x$. 
It has the two fixed points $0$ and $\infty$. 
From \eqref{eq:abc-kernel} we obtain 
\[ C(s,t) := C^H_{-1,0,1}(s,t) 
= C^H\Big(\frac{s+1}{s-1},\frac{t+1}{t-1}\Big) 
= C^H\Big(\frac{1+s}{1-s},\frac{1+t}{1-t}\Big)\] 
and 
\[ \tilde C(s,t) = 
\Big(\frac{1-s}{1+s}\Big)^H 
\Big(\frac{1-t}{1+t}\Big)^H 
C^H\Big(\frac{1+s}{1-s},\frac{1+t}{1-t}\Big).\] 
Theorem~\ref{thm:6.3} now implies that the 
kernel $\tilde C$ is reflection 
positive with respect to $(\R^\times, \R^\times_+, \theta)$. 
\end{ex}

\subsection*{Brownian bridges} 

As we shall see below, 
for $H =\shalf$, the covariance kernels $C^{1/2}_{\alpha,\gamma}$ turn out to correspond 
to Brownian bridges. 

\begin{defn} (\cite[Def.~2.8, p.~109]{Hid80}) 
(a) A Gaussian process $(X_t)_{\alpha \leq t \leq \gamma}$ is called a {\it Brownian bridge} 
if $m(t) := \E(X_t)$ is an affine function and 
\[ C(t,s) := \E((X_t - m(t))(X_s - m(s))) 
= \frac{(t \wedge s - \alpha)(\gamma - t \vee s)}{\gamma-\alpha}.\] 
If $m(t) = 0$ for every $t$, then $(X_t)_{\alpha \leq t \leq \beta}$ is called a 
{\it pinned Brownian motion}. 

(b) A {\it normalized Brownian bridge} is a 
Brownian bridge whose variance is normalized to 
$1$, so that its covariance kernels is 
\begin{equation}
  \label{eq:browbrid}
\tilde C(t,s) 
=\frac{ (s \wedge t - \alpha)(\gamma - s \vee t)}
{\sqrt{(s-\alpha)(\gamma-s)(t-\alpha)(\gamma-t)}} 
=\sqrt{\frac{ (s \wedge t-\alpha)(\gamma - s \vee t)}
{(s \vee t-\alpha)(\gamma  - s \wedge t)}}. 
\end{equation}
\end{defn} 

\begin{prop}{\rm(Reflection positivity of the Brownian bridge)} 
For $\alpha < \gamma$ in $\R$ and $H = \shalf$, the kernel 
$C^{1/2}_{\alpha,\gamma}$ is the covariance of a normalized Brownian bridge on 
the interval $[\alpha,\gamma]$. This kernel is reflection positive 
for $(X,X_+,\theta)$, where
$X = [\alpha,\gamma]$, $X_+ = [\alpha,\beta]$ 
and $\theta(t) = \alpha + \gamma-t$ is the reflection in the midpoint. 
The corresponding Hilbert space $\hat\cE$ is one-dimensional. 
\end{prop}

\begin{prf} First we observe that 
\[ C^{1/2}\Big(\frac{s-\alpha}{s-\gamma}, \frac{t-\alpha}{t-\gamma}\Big)
=  C^{1/2}\Big(\frac{s-\alpha}{\gamma-s}, \frac{t-\alpha}{\gamma-t}\Big)
= \frac{s-\alpha}{\gamma-s} \wedge \frac{t-\alpha}{\gamma-t}
= \frac{s\wedge t -\alpha}{\gamma-s \wedge t},\] 
so that we obtain for the associated normalized kernel 
\begin{align*}
C(s,t) := C^{1/2}_{\alpha,\gamma}(s,t) 
= \frac{\frac{s\wedge t -\alpha}{\gamma-s \wedge t}}
{\big(\frac{s -\alpha}{\gamma-s}\big)^{1/2}\big(\frac{t -\alpha}{\gamma-t}\big)^{1/2}}
= \sqrt{\frac{s\wedge t -\alpha}{s\vee t -\alpha}}
\sqrt{\frac{\gamma - s\vee t}{\gamma - s\wedge t}}.
\end{align*}
This is the kernel \eqref{eq:browbrid} 
of a normalized Brownian bridge on $[\alpha,\gamma]$. 

For $\beta := \frac{\alpha + \gamma}{2}$, the reflection 
$\theta^{\alpha,\beta,\gamma}$ is given by 
$\theta(t) := \alpha + \gamma - t$, 
which leaves the kernel 
$C^{1/2}_{\alpha,\gamma}$ invariant by Proposition~\ref{prop:kern-invar}. 
For $\alpha \leq t,s \leq \beta$, we have 
\[ C^\theta(s,t) = C(s, \theta(t))
= \sqrt{\frac{(s \wedge \theta(t) - \alpha)(\gamma - s \vee \theta(t))}
{(s \vee \theta(t)-\alpha)(\gamma - s \wedge \theta(t))}}
= \sqrt{\frac{(s - \alpha)(\gamma - \theta(t))}
{(\gamma - s)(\theta(t)-\alpha)}}
= \sqrt{\frac{s-\alpha}{\gamma - s}}
\sqrt{\frac{t-\alpha}{\gamma-t}}.\]
This is a positive definite kernel defining a one-dimensional 
Hilbert space. We conclude that the kernel 
$C$ is reflection positive for $(X,X_+,\theta)$, where
$X = [\alpha,\gamma]$ and $X_+ = [\alpha,\beta]$. 
\end{prf}

\section{Affine actions and fractional Brownian motion} 
\mlabel{sec:8}

In this section we discuss reflection positivity 
for the affine isometric action of $\R$ corresponding to 
fractional Brownian motion $(B_t^H)_{t \in \R}$. 
In Subsection~\ref{subsec:7.2} we shall encounter the 
curious phenomenon that, for every $H$ there exists a natural 
unitary involution $\theta$ that leads to a reflection positive 
Hilbert space, but only for $H \leq \shalf$ it can be implemented
in such a way that $\theta b^H_t = b^H_{-t}$, so that we 
obtain a reflection positive 
affine action of $(\R,\R_+,-\id)$. 
In a third subsection we discuss increments 
of a $1$-cocycle $(\beta_t)_{t \in \R}$ defining an affine isometric 
action and characterize cocycles with 
orthogonal increments as those corresponding to multiples of 
Brownian motion. 

\subsection{Generalities} 

If $(\alpha,\cE)$ with $\alpha_t \xi = U_t \xi + \beta_t$ is an affine 
isometric action of $\R$ on the complex Hilbert space $\cE$, then 
Proposition~\ref{prop:normform} implies that, up to unitary 
equivalence,  
$\cE \cong L^2(\R,\sigma)$ for a Borel measure $\sigma$ on $\R$. 
We may assume that 
\[ (U_t f)(x) = e^{itx} f(x) \quad \mbox{ and } \quad 
\beta_t(x) =e_t(x) :=\begin{cases}
\frac{e^{itx}-1}{ix} & \text{for }  x\not=0 \\   
t & \text{for }  x=0.\end{cases}\] 
Then second quantization leads to the centered Gaussian process 
$X_t := \phi(\beta_t)$ whose covariance kernel is given by 
\begin{equation}
  \label{eq:csigma}
C_\sigma(s,t) = \la e_s, e_t \ra_\cE 
= r(t) + \oline{r(s)} - r(t-s) \quad \mbox{ for } \quad 
r(t) = \int_\R \Big(1 - e^{itu} + \frac{itu}{1 + u^2}\Big)\frac{d\sigma(u)}{u^2} 
\end{equation}
(Proposition~\ref{prop:a.4b}). Below we only consider real Hilbert spaces. 
This corresponds to the situation where the measure $\sigma$ is symmetric. 
Then the function is also real and given by 
\[ r(t) = \int_\R (1 - \cos(tu))\frac{d\sigma(u)}{u^2} = \frac{1}{2} C_\sigma(t,t).\]

\begin{lem} Let $C \: \R \times \R \to \C$ be a continuous positive definite 
kernel with $C_0 = C(\cdot,0) = 0$. For $\xi \in C^\infty_c(\R)$, 
we put 
\[ C_\xi(x) := \int_\R \xi(t) C_t(x)\, dt = \int_\R C(x,t)\xi(t)\, dt.\]
Then the subspace 
\[ \{ C_\xi \: \xi \in C^\infty_c(\R)_0\}\quad \mbox{ with }\quad 
C^\infty_c(\R)_0 := \Big\{ \xi \in C^\infty_c(\R) \: \int_\R \xi(t)\, dt = 0\Big\} \] 
is dense in the corresponding 
reproducing kernel Hilbert space~$\cH_C$. 
\end{lem}

\begin{prf} From the existence of the $\cH_C$-valued integral defining $C_\xi$, 
it follows that these are elements of $\cH_C$. 
Let $\cH_1$ denote the closed subspace generated by the elements 
$C_\xi$, $\xi \in C^\infty_c(\R)_0$. 

If $\delta_n \in C^\infty_c(\R)$ is a $\delta$-sequence, 
we obtain $C_{\delta_n} \to 0$. 
For $\xi \in C^\infty_c(\R)$ with $\int_\R \xi(t)\, dt=1$, 
we have $\xi - \delta_n \in C^\infty_c(\R)_0$, so that 
$C_\xi - C_{\delta_n} \in \cH_1$  and $C_{\delta_n} \to 0$ imply that 
$C_\xi \in \cH_1$. 
Using a sequences of the form $\xi_n := \delta_n(\cdot - t) \in C^\infty_c(\R)$, which 
converges to $\delta_t$, we see that $C_{\xi_n} \to C_t$ for $t \in \R$, 
hence that $\cH_1 = \cH_C$. 
\end{prf}

In the following we write $\cS(\R)_0 := \big\{ \xi \in \cS(\R) \: \int_\R \xi = 0\big\}$. 

\begin{cor} Let $\alpha_t \xi = U_t \xi + \beta_t$ define a 
continuous affine isometric action of $\R$ on the real Hilbert space $\cH$. 
For $\xi \in \cS(\R)$ we put $\beta_\xi := \int_\R \xi(t)\beta_t\, dt$. 
Then $\{\beta_\xi \: \xi \in \cS(\R)_0 \}$  generates the same closed 
subspace as $(\beta_t)_{t \in \R}$. 
\end{cor}

\begin{rem} \mlabel{rem:newscal-schwartz}
(a) A function $D \: \R \to \R$ with $D(0) = 0$ 
is negative definite if and only if 
\begin{equation}
  \label{eq:cd-rel}
 C(s,t) := \frac{1}{2}(D(s) + D(t) - D(s-t))
\end{equation}
is a positive definite kernel 
(Lemma~\ref{lem:bcr}).
Then $D(s) = C(s,s)$ yields 
\[ D(s-t) = C(s,s) + C(t,t) -2 C(s,t)\]  
On $\cS(\R)_0$ we therefore obtain with \eqref{eq:cd-rel} 
\begin{equation}
  \label{eq:C-D-rel}
-\frac{1}{2}\int_\R \oline{\xi(x)} \eta(y) D(x-y)\, dx\, dy 
= \int_\R \oline{\xi(x)} \eta(y) C(x,y)\, dx\, dy \quad \mbox{ for } \quad 
\xi,\eta \in \cS(\R)_0
\end{equation}
(cf.\ \cite[p.~222]{Ga67} for a corresponding statement on 
more general homogeneous spaces). 

(b) Write $C(s,t) = \la \beta_s,\beta_t\ra$ and 
$D(t) = C(t,t)$ for a 
cocycle $(\beta_t)_{t \in \R}$ of an orthogonal representation 
$(U_t)_{t \in \R}$ of $\R$ on the real Hilbert space~$\cE$ 
(see Section~\ref{sec:4} and \cite{Gu72, HV89}).
We also assume that the family $(\beta_t)_{t \in \R}$ is total in $\cE$. 
Then 
\[ \la \beta_\xi, \beta_\eta \ra_{\cE} 
= \int_\R \int_\R \oline{\xi(s)}\eta(t)C(s,t)\, dt\, ds
= -\frac{1}{2}\int_\R \int_\R \oline{\xi(s)}\eta(t)D(s-t)\, dt\, ds
\quad \mbox{ for }\quad \xi,\eta \in \cS(\R)_0.\] 

As $\cE$ is generated by the $(\beta_t)_{t \in \R}$, the Hilbert space 
$\cE$ can be identified with the reproducing kernel Hilbert space 
$\cH_C \subeq \cS'(\R)$ corresponding to the positive definite 
distribution~$C$, but the preceding argument adds another picture. 
It can also be identified with the Hilbert space 
$\cH_D$ obtained by completing $\cS(\R)_0$ with respect to the 
scalar product \eqref{eq:C-D-rel}. Taking into account that 
$\cS(\R)_0 = \{ \xi' \: \xi \in \cS(\R)\}$, this is how 
we introduced the Hilbert space $\cH_H$ in \eqref{eq:newscal}. 
\end{rem}

\subsection{Fractional Brownian motion} 
\mlabel{subsec:7.2}

For fractional Brownian motion with Hurst index $H \in (0,1)$, we have 
\[ C^H(s,t) = \frac{1}{2}(|s|^{2H} + |t|^{2H} - |s-t|^{2H}) 
\quad \mbox{ and } \quad  D^H(t) = C^H(t,t) = |t|^{2H}.\]
As in \eqref{eq:csigma}, the spectral measure $\sigma$ (a Borel measure on $\R$) 
of fractional Brownian motion is determined by 
\[ C^H(s,t) = \int_\R \oline{e_s(\lambda)} e_t(\lambda)\, d\sigma(\lambda) 
\quad \mbox{ for } \quad 
e_t(\lambda) =
\begin{cases}
\frac{e^{i\lambda t}-1}{i\lambda} & \text{ for } \lambda\not=0 \\   
t & \text{ for }  \lambda=0.\end{cases}\] 
The corresponding realization is obtained by 
$\beta_t := e_t \in L^2(\R,\sigma) =: \cE$ 
(Proposition~\ref{prop:normform}). 
According to \cite[p.~40]{AJP12}, the measure $\sigma$ 
is given by 
\[ d\sigma(\lambda) = \frac{\sin(\pi H) \Gamma(1 + 2 H)}{2\pi} \cdot 
|\lambda|^{1-2H}\, d\lambda\] 
(see \eqref{eq:e.1} in Appendix~\ref{app:e} for a derivation of this formula). 

\begin{ex} For $H = \shalf$, the spectral measure $\sigma$ is a 
multiple of Lebesgue measure: 
\[ d\sigma(\lambda) 
= \frac{\sin(\pi/2) \Gamma(2)}{2\pi} \, d\lambda
= \frac{d\lambda}{2\pi}.\] 

This leads to a 
natural realization of Brownian motion 
by a cocycle of the multiplication representation 
of $\R$ on $L^2(\R)$ and, by Fourier transform, to the realization 
of Brownian motion as a cocycle for the translation representation 
of $\R$ on $L^2(\R)$. 
\end{ex}

\begin{rem} Combining Remark~\ref{rem:newscal-schwartz} with 
\eqref{eq:newscal} in Section~\ref{sec:5}, we see that the Hilbert space $\cH_H$ can alternatively 
be constructed from the scalar product 
\[ \la \xi,\eta \ra_{C^H}  
= -\frac{1}{2} \int_\R \int_\R \oline{\xi(x)}\eta(y) |x-y|^{2H}\, dx\, dy 
= \int_\R \int_\R \oline{\xi(x)}\eta(y) C^H(x,y)\, dx\, dy 
\quad \mbox{ for } \quad \xi,\eta \in \cS(\R)_0 \]
as the completion of $\cS(\R)_0$. 
This implies that the map 
\[ D \: \cS(\R) \to \cS(\R)_0, \quad \xi \mapsto \xi' 
\quad \mbox{ satisfies } \quad 
\|D\xi\|_{C^H} = \|\xi\|_H,\] 
so that $D$ extends to a unitary operator 
$D \:  \cH_H \to \cH_{C^H}$. Here we write $\cH_{C^H} \subeq \cS'(\R)$ 
for the Hilbert space of distributions defined by $C^H$, obtained by 
completing $\cS(\R)$ with respect to the scalar product $\la \cdot,\cdot \ra_{C^H}$, 
defined by the positive definite distribution kernel~$C^H$ 
and associating to $\xi \in \cS(\R)_0$ the distribution 
$C_\xi^H := \la \cdot, \xi \ra_{C^H}$.
Note that 
\[ D(b^H_t)  
= D(\chi_{[0,\infty)} - \chi_{[t,\infty)})
= \delta_0 - \delta_t,\] 
where we consider the $\delta$-functionals as elements of $\cH_{C^H}$ with
\[ \la \delta_s, \xi \ra_{C^H}  
= \int_\R \int_\R \delta_s(x) \xi(y) C^H(x,y)\, dx\, dy 
=  \int_\R C^H(s,y)\xi(y) \, dy =  C^H_\xi(s).\] 
As a distribution, $\delta_s$ corresponds to the function $C^H_s$ which 
corresponds to the evaluation in~$s$ in the reproducing kernel 
Hilbert space $\cH_{C^H}\subeq \cS'(\R)$. Compare also with the 
corresponding discussion in \cite{NO14} and \cite[Ch.~7]{NO18}. 

The inverse of the  unitary operator $D \: \cH_H \to \cH_{C^H}$ is given by 
\[ I \:  \cH_{C^H} \to \cH_H, \quad 
I(\xi)(x) 
= \int_{-\infty}^x \xi(y)\, dy = - \int_x^{\infty} \xi(y)\, dy 
\quad \mbox{ for } \quad \xi \in \cS(\R)_0.\]  
\end{rem}
 
\begin{prop} Consider the realization $(b^H_t)_{t \in \R}$  
of fractional Brownian motion in the 
Hilbert space $\cE := \cH_H$, 
the affine isometric $\R$-action defined by 
\[ \alpha_t^H \xi = S_t^H \xi + b^H_t, \] 
where $S_t^H$ denotes the translation by $t$ on $\cH_H$, 
and the closed subspace $\cE_+$ generated by $(b^H_t)_{t \geq 0}$. 
Then 
\[ (\theta\xi)(x) := -\xi(-x) \] 
defines a unitary involution with 
$\theta b^H_t = b^H_{-t}$ for $t \in \R$. 
Now $(\cE,\cE_+,\theta)$ is reflection positive if and only if $H \leq \shalf$, 
so that we obtain in this case a reflection positive affine action. 
For $H \geq \shalf$, the triple $(\cE,\cE_+,-\theta)$ is also 
reflection positive, but it does not lead to a reflection positive affine 
action because $-\theta b_t^H \not= -b_t^H \not= b_t^H$ for $t > 0$. 
\end{prop}

\begin{prf} By Example~\ref{ex:repo-realline}(a), 
the function $D^H(t) = |t|^{2H}$ is negative definite on the additive group $(\R,+)$ 
and it is reflection positive for $(\R,\R_+,-\id)$ 
if and only if $H \leq 1/2$ by Example~\ref{ex:repo-realline}(b). 
Accordingly, the reflection 
$\sigma(t) = -t$ on $\R$ leads to the twisted kernel 
\[ C^H(s,-t) = \frac{1}{2}(|s|^{2H} + |t|^{2H} - |t+s|^{2H}) \] 
on $\R_+$ which is positive definite if and only if $D^H$ is negative 
definite on $(\R_+, \id)$ (Lemma~\ref{lem:bcr}), 
which in turn is equivalent to $H \leq \shalf$.  

For $H \geq \shalf$, the unitary involution $-\theta$ satisfies 
$-\theta b^H_t = - b^H_{-t}$, 
which leads to the twisted kernel 
\[ (s,t) \mapsto 
\la b^H_s, -\theta b^H_{t} \ra 
= \la b^H_s, -b^H_{-t} \ra = -C^H(s,-t)
= -\frac{1}{2}(|s|^{2H} + |t|^{2H} - |t+s|^{2H}). \] 
As $-D^H(t) = -|t|^{2H}$ is negative definite on the semigroup $(\R_+, \id)$ 
if $\shalf \leq H \leq 1$ (Example~\ref{ex:repo-realline}(c)), the assertion follows.
\end{prf}

We conclude that the affine actions of $\R$ corresponding to 
fractional Brownian motion with Hurst parameter $H \leq \shalf$ 
leads to a reflection positive affine action, and 
from the calculation in Example~\ref{ex:7.2} we derive that 
the reflection positive function on 
$(\R,\R_+, -\id)$ corresponding to the constant function 
$1 = e^{i\phi(0)} \in \Gamma(\cE)$ is given by 
$\vphi(t) = e^{-\|b^H_t\|^2/2} = e^{-t^{2H}/2}$. 
The preceding proposition also explains why we obtain trivial reflection positivity 
for $H = \shalf$ since in this case 
$(\cE,\cE_+, \pm\theta)$ are both reflection positive. 

\begin{rem} For $h > 0$ and $H \leq \shalf$ and the kernel 
$Q^h(x,y) := e^{-h\|x-y\|^2}$ 
on $\cE$, we obtain with the same arguments the positive definite functions 
$\vphi(t) = e^{-h t^{2H}}$ on $\R_+$. 
\end{rem}

\subsection{Cocycles with orthogonal increments}

In this subsection we discuss the question when a cocycle 
$(\beta_t)_{t \in \R}$ for an orthogonal representation 
$(U,\cH)$ of $\R$ has {\it orthogonal increments} in the sense that, 
for $t_1 \leq t_2 \leq t_3 \leq t_4$ we have
 \[ \la \beta_{t_2} - \beta_{t_1}, \beta_{t_4} - \beta_{t_3} \ra = 0\] 
(cf.\ \cite{Ko40a}). 

 \begin{prop} \mlabel{prop:ortho-inc} The following are equivalent: 
   \begin{itemize}
   \item[\rm(i)] $\beta$ has orthogonal increments.
   \item[\rm(ii)] For $s,t > 0$, we have $\la \beta_t, \beta_{-s} \ra = 0$. 
   \item[\rm(iii)] There exists a $c \geq 0$ such that 
$C(t,s) := \la \beta_t, \beta_s \ra =  \frac{c}{2}(|s| + |t| - |s-t|)$ 
for all $t,s \in \R$. If $c > 0$, then 
$(c^{-1/2}\phi(\beta_t))_{t \in \R}$ realizes a two-sided Brownian motion in $\Gamma(\cH)$. 
   \end{itemize}
 \end{prop}

 \begin{prf} (cf.~\cite[Satz~8]{Ko40a} for a variant of this observation) 

(i) $\Rarrow$ (ii) follows with 
$t_1 = - s, t_2 = t_3 = 0$ and $t_4 = t$. 

(ii) $\Rarrow$ (i): Let $\cH_\pm \subeq \cH$ be the closed subspaces 
generated by the $\beta_t$ for $\pm t \geq 0$. Then (ii) means that 
$\cH_+ \bot \cH_-$. For $t_1 \leq t_2 \leq t_3 \leq t_4$ we now observe that 
\[ \beta_{t_1} -\beta_{t_2}  
=  \beta_{t_2 + (t_1 - t_2)} - \beta_{t_2} 
=  U_{t_2} \beta_{t_1 - t_2} \] 
and similary  
$\beta_{t_4} -\beta_{t_3}  =  U_{t_3} \beta_{t_4-t_3}.$ 
Therefore 
\begin{align*}
\la \beta_{t_1} - \beta_{t_2}, \beta_{t_4} - \beta_{t_3} \ra 
&= \la U_{t_2} \beta_{t_1 - t_2}, U_{t_3} \beta_{t_4 - t_3} \ra
= \la \beta_{t_1 - t_2}, U_{t_3-t_2} \beta_{t_4 - t_3} \ra \\
&= \la \beta_{t_1 - t_2}, \beta_{t_3 - t_2 + t_4 - t_3} - \beta_{t_3-t_2} \ra=0.
\end{align*}

(ii) $\Rarrow$ (iii): Put $\psi(t):= C(t,t) = \|\beta_t\|^2$ and note that this 
function is increasing for $t \geq 0$. 
For $0 \leq s \leq t$ we have 
\[ C(s,t) = \la \beta_s, \beta_t \ra 
= \la \beta_s, (\beta_t - \beta_s) + \beta_s \ra 
= \la \beta_s, \beta_s \ra = \psi(s),\] 
and therefore 
\begin{equation}\label{eq:c-psi-rel}
C(s,t) = \psi(s \wedge t)\quad \mbox{ for } \quad t,s \geq 0.
\end{equation}
Further,
\[ \|\beta_t - \beta_s\|^2 
= C(t,t) + C(s,s) - 2 C(t,s)
= \psi(t) + \psi(s) - 2 \psi(s \wedge t) 
= |\psi(t) - \psi(s)|,\] 
so that translation invariance of this kernel leads to 
\[ \|\beta_t - \beta_s\|^2 = \psi(|t-s|).\] 
From the orthogonality of the increments, we further derive for 
$0 \leq s \leq t$ the relation 
\[ \psi(t) - \psi(s) = \psi(t-s),\] 
so that 
\[ \psi(a + b) = \psi(a) + \psi(b) \quad \mbox{ for } \quad a,b \geq 0.\] 
Since $\psi$ is continuous, there exists a $c\geq 0$ with 
$\psi(t) = ct$ for $t \geq 0$, and therefore \eqref{eq:c-psi-rel} yields 
$C(s,t) = c \cdot s \wedge t$ for $t,s \geq 0$. 

We likewise find some $c' \geq 0$ with 
\[ C(s,t) = c' \cdot |s| \wedge |t|  \quad \mbox{ for } \quad s,t \leq 0.\] 
Now $\psi(-t) = \psi(t)$ implies that $c' = c$, and this completes the proof. 

(iii) $\Rarrow$ (ii) follows from the fact that $C(s,t) = 0$ for $ts < 0$ holds 
for the covariance kernel of Brownian motion. 
\end{prf}

If $(\cE,\cE_+, \theta)$ is reflection positive 
for an affine $\R$-action,  $\cE_+$ is generated by $(\beta_t)_{t \geq 0}$ 
and $\theta \beta_t = \beta_{-t}$, 
then the space $\hat\cE$ is trivial if and only if 
\[ \la \beta_s, \beta_t \ra = 0 \quad \mbox{ for } \quad ts < 0,\] 
which in turn means that $\beta$ has orthogonal increments 
by Proposition~\ref{prop:ortho-inc}. 
In view pf Proposition~\ref{prop:ortho-inc}(iii),  
Brownian motion can, up to positive multiples, 
be characterized as a process with stationary orthogonal increments. 

\begin{rem} Consider the stochastic process $(\phi(\beta_t))_{t \in \R}$ 
associated to the cocycle $(\beta_t)_{t \in \R}$ in $\cE$. 
We say that the {\it increments} of this process are 
{\it positively (negatively) correlated} if, 
for $t_1 \leq t_2 \leq t_3 \leq t_4$, we have
\[ \pm \la \beta_{t_2} - \beta_{t_1}, \beta_{t_4} - \beta_{t_3} \ra \geq 0. \] 
As 
\begin{align*}
\la \beta_{t_2} - \beta_{t_1}, \beta_{t_4} - \beta_{t_3} \ra 
= \la U_{t_1} \beta_{t_2 - t_1}, U_{t_3} \beta_{t_4 - t_3} \ra 
= \la U_{t_1- t_3} \beta_{t_2 - t_1}, \beta_{t_4 - t_3} \ra
= \la \beta_{t_2-t_3} - \beta_{t_1- t_3}, \beta_{t_4 - t_3} \ra, \end{align*}
we may w.l.o.g.\ assume that $t_3 = 0$, i.e., 
$t_1 \leq t_2 \leq 0 \leq t_4$. Therefore 
the process has positively (negatively) correlated increments if and only if, for 
every $t \geq 0$, the functions 
\[  C_t(s) := C(s,t) := \la \beta_s, \beta_t \ra \] 
are increasing (decreasing) on $(-\infty, 0]$. Note that this implies in 
particular that $C(s,t) \geq 0$, resp., $\leq 0$ for $s \leq 0 \leq t$. 

For fractional Brownian motion, we have for $s \leq 0 \leq t$ 
\[ C^H(s,t) = \frac{1}{2}((-s)^{2H} + t^{2H} - (t-s)^{2H}).\] 
As 
\[ \frac{\partial}{\partial s} C^H(s,t) = H((t-s)^{2H-1}-(-s)^{2H-1}) \]
is non-negative for $H \geq \shalf$ and non-positive for $H \leq \shalf$, 
it follows that fractional Brownian motion has positively 
correlated increments for $H \geq \shalf$ and 
negatively correlated increments for $H \leq \shalf$. 
\end{rem}

\section{Perspectives} 
\mlabel{sec:9}

In this final section we briefly discuss some results 
that are possibly related to far reaching generalizations 
of what we discuss in the present paper on the real line, resp., 
on its conformal compactification~$\bS^1$.

\subsection{Helices and Hilbert distances} 

Let $G$ be a Lie group, $K\subeq G$ be a closed subgroup. 
We write $X = G/K$ for the corresponding homogeneous space  
and $x_0 := eK$ for the base point in~$X$.

In \cite[Def.~2.3]{Ga67}, a kernel $C \: X \times X \to \R$ 
on $X = G/K$ is called a {\it L\'evy--Schoenberg kernel} if 
\begin{itemize}
\item[\rm(LS1)] $C$ is positive definite and $C_{x_0} = 0$.  
\item[\rm(LS2)] The kernel $r(s,t) := C(s,s) + C(t,t) - 2 C(s,t)$ is 
$G$-invariant. 
\end{itemize}
Then $\psi(g) := r(x_0, x) = C(x,x)$, where $x = g.x_0$, 
defines a function on $G$ with 
\begin{equation}
  \label{eq:levyker}
C(gK,hK) =  \frac{1}{2}(\psi(g) + \psi(h) - \psi(h^{-1}g)),
\end{equation}
so that $\psi$ is a negative definite $K$-biinvariant function with 
$\psi(e) = 0$ on $G$ (Lemma~\ref{lem:bcr}).
Conversely, every such function defines by \eqref{eq:levyker} a 
L\'evy--Schoenberg kernel on $G/K$. 

For a L\'evy--Schoenberg kernel  $C$, there exists a 
map $\xi \: X \to \cE$ into a real Hilbert space $\cE$ 
with $\xi(x_0)= 0$, unique up to orthogonal equivalence (Lemma~\ref{lem:ex-isom}), 
such that 
\[ r(x,y) = \|\xi(x)-\xi(y)\|^2 \quad \mbox{ for } \quad x,y \in X.\] 
Then $\xi$ is called a {\it helix} and $\sqrt{r}$ is called an 
{\it invariant Hilbert distance} on~$X$. The uniqueness of $\xi$ 
further implies the existence of an affine isometric action 
$\alpha \: G \to \Mot(\cE)$ for which $\xi$ is equivariant. 
Writing $\alpha_g\xi = U_g\xi + \beta_g$, we then have 
$\xi(gK) = \beta_g$ for $g \in G$. 
In particular, any helix specifies an orthogonal representation~$(U,\cE)$ of~$G$. 

Classification results for L\'evy--Schoenberg kernels, resp., 
invariant Hilbert distances, resp., affine isometric actions 
of $G$ with a $K$-fixed points, are mostly stated 
in terms of integral formulas (L\'evy--Khintchine formulas). 
Results are nown in various contexts: 
\begin{itemize}
\item for $G$ locally compact and $K$ compact (\cite{FH74}); 
see \cite{PRV63} and \cite{Ha69} for locally compact abelian groups. 
\item for the euclidean motion group $G = E(d) \supeq \OO_d(\R) = K$ 
(\cite{vNS41}, \cite[p.~135]{Ga67})
\item for $G$ compact (\cite[Thm.~3.15]{Ga67}); see \cite{Bo41} 
for $G = \SO_{d+1}(\R)$ and $X = \bS^d$. 
\item for $G/K$ Riemannian symmetric (\cite[Thm.~3.31]{Ga67} and 
\cite[Thm.~4.1]{Ga67} for $G = \SL_2(\R)$)
\item for $G$ the additive group of a Hilbert space and $K$ a closed subspace 
(\cite{Va62}). 
\item for $G = \OO_{1,\infty}(\R)$ and $K = \OO_\infty(\R)$ and 
$X$ the infinite dimensional hyberbolic space (\cite[Thm.~8.1]{FH74}). 
It is shown in particular that the kernel $Q(x,y) = \log\cosh(d(x,y))$ is 
negative definite, so that all kernels $e^{-s Q(x,y)} = \cosh(d(x,y))^{-s}$ 
are positive definite; they correspond to the spherical 
functions of~$X$ (cf.\ \cite[Thm.~21, p.~79]{HV89}).
\end{itemize}

\begin{exs} \mlabel{exs:8.1} (a) If $\cE$ is a real Hilbert space 
and $G = \Mot(\cE) \cong \cE \rtimes \OO(\cE)$ its isometry group, 
then $\psi(b,g) := \|b\|^{2H}$ defines for $0 < H \leq 1$ a 
negative definite $\OO(\cE)$-biinvariant function on $G$ with $\psi(e) = 0$ 
(\cite[Ex.~3.2.13(b)]{BCR84}). 
The corresponding L\'evy--Schoenberg kernel on $\cE$ is 
\[ C(s,t) := \frac{1}{2}(\|s\|^{2H} +\|t\|^{2H} - \|s-t\|^{2H}) 
\quad \mbox{ with } \quad r(s,t) = \|s-t\|^{2H} \] 
(cf.\ \cite[p.~135]{Ga67}). 

(b) The kernel 
\begin{equation}
  \label{eq:metric}
 C(s,t) := \frac{1}{2}(d(s,e_0)+ d(t,e_0)- d(s,t))
\quad \mbox{ with }\quad r(s,t) = d(s,t) 
\end{equation}
on the sphere $\bS^d$, where $d$ denotes the Riemannian metric on $\bS^d$ 
and $e_0 \in \bS^d$ is fixed. Here $G = \OO_{d+1}(\R)$ and $K = \OO_d(\R)$ 
is the stabilizer of~$e_0$ (\cite[p.~174]{Ga67}, \cite{Le65}). 
This means that the Riemannian metric on $\bS^d$ is a negative definite kernel.

(c) From (a) it follows that for any real-valued negative definite function 
$\psi$ satisfying $\psi(e) = 0$ on the group $G$, the functions $\psi^H$, $0 \leq H \leq 1$, 
are negative definite as well (\cite[p.~189]{Ga67}). 
\end{exs}

\subsection{Brownian motion on metric spaces} 

\begin{defn} Let $(M,d)$ be a metric space. 
In \cite{Le65} a real-valued Gaussian process $(B_m)_{m \in M}$ 
is called a {\it Brownian motion with parameter space $(M,d)$} 
if there exists a point $m_0 \in M$ with 
\[ \bE(B_n B_m)= \frac{1}{2}\big(d(m,m_0)+ d(n,m_0) - d(m,n)\big)\quad \mbox{ for }\quad 
m,n \in M.\] 
\end{defn}

\begin{rem} For a metric space $(M,d)$ a Brownian motion with parameter space 
$(M,d)$ exists if and only if the metric $d \: M \times M \to \R$ is a negative 
definite kernel, which is equivalent to the kernels 
\[ C(n,m) = \frac{1}{2}\big(d(m,m_0)+ d(n,m_0) - d(m,n)\big) \] 
being positive definite for every $m_0 \in M$ 
(Lemma~\ref{lem:bcr}; see also \cite{Ta77} and \cite[Cor.~58]{Le65}). 
This is verified for $M = \R^d$ in \cite[Thm.~7]{Ta77} and for 
$M = \bS^d$ in \cite[Thm.~5]{Ta77}. 
\end{rem} 

\begin{defn}
If $d$ is negative definite, then there exists an isometric embedding 
$\eta \: M \to \cE$ into a real Hilbert space with $\eta(m_0) = 0$ and then 
\[ C(n,m) = \frac{1}{2}\big(\|\eta(m)\| + \|\eta(n)\| - \|\eta(m)-\eta(n)\|\big).\] 
Example~\ref{exs:8.1}(a) then implies that the kernels 
\[ C^H(n,m) = \frac{1}{2}\big(\|\eta(m)\|^{2H} + \|\eta(n)\|^{2H} - 
\|\eta(m)-\eta(n)\|^{2H}\big), \qquad 0 \leq  H \leq  1,\] 
are positive definite as well. This suggests to call a Gaussian process 
$(B_m^H)_{m \in M}$ a {\it fractional Brownian motion with parameter space $(M,d)$ 
and Hurst index $H \in (0,1)$} if there exists an $m_0 \in M$ with 
$\bE(B_n^H B_m^H)= C^H(n,m)$ for $n,m \in M$. 
Note that $(\phi(\eta(m)))_{m \in M}$ yields a realization of fractional Brownian 
motion with parameter space $(M,d)$ in the Fock space $\Gamma(\cE)$.
If $M = G/K$ is a homogeneous space, then the map 
$\eta$ is called a {\it fractional Brownian helix} (cf.~\cite{Ka81} for the 
terminology). 

\end{defn}

For various aspects of fractional Brownian motion on $\R^d$, we refer to 
\cite{AJ12} and \cite{AJL11}.

\begin{prob} The natural analog of the function $\chi_{\R_+}$ which generates 
the realization of the fractional Brownian motion in $\cH_H$ has a 
natural higher dimensional analog in $\chi_{\R^d_+}$, the characteristic function 
of a half space. Does this correspond to some ``fractional Brownian motion'' on $\R^d$? 
\end{prob}

\subsection{Complementary series of the conformal group} 

The function $\|x\|^{-\alpha}$ on $\R^d$ is locally integrable if and only if 
$\alpha < d$, and it defines a positive definite distribution if and only if 
$\alpha \geq 0$ (\cite[Lemma~2.13]{NO14}). 
We thus obtain a family of Hilbert subspaces 
$\cH_\alpha \subeq C^{-\infty}(\R^d)$ for $0 \leq \alpha < d$. 
For $\alpha = 0$ this space is one-dimensional, consisting of constant functions. 

From \cite[Prop.~6.1]{NO14} we also know that, for $0 \leq \alpha < d$, the distribution 
$\|x\|^{-\alpha}$ is reflection positive with respect to 
$\theta(x) = (-x_0, \bx)$ if and only if $\alpha = 0$ or $d-2 \leq \alpha < d$. 

Let $G := \Conf(\R^d) \subeq \Diff(\bS^d)$ be the conformal group of $\R^d$, 
considered as a group of diffeomorphisms of the conformal 
compactification $\bS^d$ (implemented by a stereographic projection). 
We consider the kernels 
\[ Q(x,y) := \|x-y\|\quad \mbox{ and } \quad 
Q_\alpha(x,y) := \|x-y\|^{-\alpha}.\] 
We then have 
\begin{equation}
  \label{eq:q-trafo}
Q(g(x),g(y)) = \|\dd g(x)\|^{1/2} Q(x,y) \|\dd g(y)\|^{1/2} \quad \mbox{ if } \quad 
g(x),g(y)\in \R^d.
\end{equation}
In fact, this relation is obvious for affine maps $g(x) = Ax + b$, 
$A \in \R^\times \OO_d(\R)$. As the conformal group is generated by the 
affine conformal group $\R^d \rtimes (\R^\times\OO_d(\R))$ and the 
inversion $\sigma(x) := \frac{x}{\|x\|^2}$ in the unit sphere, 
it now suffices to verify the relation also for $\sigma$. It is a consequence 
of 
\begin{align*}
 \|\sigma(x)-\sigma(y)\|^2 
&= \Big\|\frac{x}{\|x\|^2} - \frac{y}{\|y\|^2} \Big\|^2 
= \frac{1}{\|x\|^{2}\|y\|^{2}} \Big\| \frac{\|y\| x}{\|x\|} -\frac{ \|x\|y}{\|y\|}\Big\|^2 \\
&= \frac{1}{\|x\|^{2}\|y\|^{2}}(\|y\|^2 - 2 \la x, y \ra + \|x\|^2) 
=\frac{\|y-x\|^2}{\|x\|^{2}\|y\|^{2}},
\end{align*}
combined with 
\[ \dd\sigma(x)y 
= \frac{y}{\|x\|^2} - 2 \frac{\la x, y \ra}{\|x\|^4}x 
= \frac{r_x(y)}{\|x\|^{2}},\] 
where $r_x$ is the reflection in $x^\bot$, so that $\|d\sigma(x)\| = \|x\|^{-2}$. 

As a consequence, we obtain 
\begin{equation}
  \label{eq:alpha-trafo}
Q_\alpha(g.x, g.y) = \|\dd g(x)\|^{-\alpha/2} Q_\alpha (x,y) 
\|\dd g(y)\|^{-\alpha/2} 
\end{equation}
(see \cite[Lemma~5.8]{NO14} for the corresponding relation on the sphere $\bS^d$). 

The transformation formula \eqref{eq:q-trafo} 
implies in particular that the {\it conformal cross ratio} 
\[ \CR(x, y,z,u)
:= \frac{Q(x,y) Q(z,u)}
{Q(x,u) Q(z, y)} \] 
is invariant under the conformal group. 

In view of \eqref{eq:alpha-trafo}, we obtain with $J_g(x) := \|\dd g(x)\|$ a representation 
\[ (U^\alpha_g \xi)(x) 
:= J_{g^{-1}}(x)^{d-\frac{\alpha}{2}} \xi(g^{-1}.x)  
\] 
on test functions. The same calculation as in \cite[Lemma~5.8]{NO14} 
now implies that $U^\alpha$ defines a unitary representation of $G$ 
on the space $\cH_\alpha$, specified by the scalar product 
\[ \la \xi,\eta \ra_\alpha 
:= \int_{\R^d} \int_{\R^d} \oline{\xi(x)} \eta(y) Q_\alpha(x,y)\, dx\, dy
= \int_{\R^d} \int_{\R^d} \oline{\xi(x)} \eta(y) 
\|x-y\|^{-\alpha}
\, dx\, dy \quad \mbox{ for } \quad \xi,\eta \in \cS(\R^d).\] 
For $g^{-1}(x) = Ax + b$, we have in particular 
\[ (U^\alpha_{g^{-1}} \xi)(x) 
= \|A\|^{d-\frac{\alpha}{2}} \xi(Ax+ b),\] 
and for the involution $\sigma(x) = \|x\|^{-2}x$ we have 
\[ (U^\alpha_{\sigma} \xi)(x) 
= \|x\|^{\alpha - 2d}\xi(\sigma(x)).\]

\begin{rem}
Up to the factor $\sgn(\det g)$, this specializes for $d = 1$ and 
$\alpha = 2(1-H)$ to the representation $U^H$ for 
$\shalf< H < 1$. We refer to Appendix \ref{app:d} for more detailed discussion
of this case.
\end{rem}

As the kernel $D^H(x,y) := \|x-y\|^{2H}$, $0 < H \leq 1$, 
on $\R^d$ is negative definite, the corresponding kernel 
\[  C^H(x,y) := \frac{1}{2}\big(\|x\|^{2H} + \|y\|^{2H}- \|x-y\|^{2H}\big) \] 
is positive definite. We thus obtain on 
$\cS(\R^d)_0 = \big\{ \xi \in \cS(\R^d) \: \int_{\R^d} \xi(x)\, dx = 0 \big\}$ 
a positive semidefinite hermitian form by 
\begin{align}
  \label{eq:8.3}
\la \xi,\eta\ra_{C^H} 
&:= \int_{\R^d} \int_{\R^d} 
\oline{\xi(x)}\eta(y) C^H(x,y)\, dx\, dy \nonumber\\
&= - \frac{1}{2} \int_{\R^d} \int_{\R^d} 
\oline{\xi(x)}\eta(y) \|x-y\|^{2H}\, dx\, dy, \qquad  \xi,\eta \in \cS(\R)_0.
\end{align}

\begin{ex} For $H =1$, we have 
\[  C^1(x,y) = \frac{1}{2}\big(\|x\|^2 + \|y\|^2- \|x-y\|^{2}\big)
= \la x, y\ra,\] 
so that the corresponding reproducing kernel Hilber space is $\cH_{C^1} \cong \C^d$. 
For $\xi,\eta \in \cS(\R^d)$, we then have 
\[ \la \xi,\eta\ra_{C^1} 
=\la [\xi], [\eta] \ra, 
\quad \mbox{ where } \quad [\xi] := \int_{\R^d} \xi(x) x\, dx \in \R^d\] 
is the {\it center of mass} of the measure $\xi\, dx$. 
\end{ex}

\begin{rem} In \cite[Thm.~7]{Ta90} Takenaga derives some ``conformal invariance'' 
of Brownian motion in $\R^d$ but it seems that his method only 
works on the parabolic subgroups of the conformal group stabilizing 
either $0$ or $\infty$. So it would be interesting to use 
the complementary series representations of the conformal group to 
derive a more complete conformal invariance in the spirit of the present 
paper for $d > 1$. 
\end{rem}

\begin{rem} Similar arguments as in Remark~\ref{rem:5.x} 
apply in the higher dimensional context: 
Since the complementary series 
representations $(U^\alpha, \cH_\alpha)$ of $\OO_{1,d}(\R)^\uparrow$ are irreducible, 
\cite[Prop.~5.20]{JNO16a} implies that the space 
$\cH_\alpha^\infty$ of smooth vectors is nuclear. 
From the proof of \cite[Prop.~5.20(b)]{JNO16a}, we further derive that 
an element $\xi \in \cH_\alpha$ is a smooth vector if and only if it is a smooth 
vector for the maximal compact subgroup $K \cong \OO_d(\R)$. Considering 
$\cH_\alpha$ as a space of distributions on the sphere $\bS^d$, it is not hard to see 
that $\cH_\alpha^\infty = C^\infty(\bS^d)$ and hence that 
$\cH_\alpha^{-\infty} = C^{-\infty}(\bS^d)$ is the space of distributions on the sphere. 
\end{rem}

\subsection{The Ornstein--Uhlenbeck process} 
 
In this section we describe shortly the connection to the {\it Ornstein--Uhlenbeck
process}. For that let  $H = \frac{1}{2}$. Then  
$Y_t := \phi(\tau^{1/2}_{e^t}\chi_{[0,1]})$, $t \in \R$, is
a stationary Gaussian process realized in $\cH_{1/2} \cong L^2(\R)$. It is 
the {\it Ornstein--Uhlenbeck process}.  The corresponding covariance kernel is 
\[ C(t,s) := \E(Y_t Y_s) = \int_0^{e^{s-t}} e^{(t-s)/2}\, du = e^{(s-t)/2} 
= e^{-|s-t|/2} 
 \quad \mbox{ for } \quad s \leq t\]
 which is reflection positive with respect to 
$(\R, \R_+, -\id_\R)$ because the kernel 
\[ C(-t,s) = e^{-(s+t)/2} \] 
on $\R_+$ is positive definite leading to a one-dimensional Hilbert space
via the Osterwalder-Schrader construction (cf.~Example~\ref{ex:1.3}).

For $0 < H < 1$, we also obtain by 
$Y_t^H := \phi(\tau^H_{e^t}\chi_{[0,1]})
= \phi(e^{tH} \chi_{[0,e^{-t}]})$, $t \in \R$, 
in $\cH_H$ a stationary Gaussian process. 
The corresponding covariance kernel is 
\begin{align*}
C(t,s) 
&= \E(Y_s Y_t) 
= e^{(t+s)H} C^H(e^{-s}, e^{-t}) 
= \frac{e^{(t+s)H}}{2}(e^{-2sH} + e^{-2tH} - |e^{-t} - e^{-s}|^{2H}) \\
&= \frac{1}{2}\big(e^{(t-s)H} + e^{(s-t)H} - 
|e^{(s-t)/2} - e^{(t-s)/2}|^{2H}\big) \\
&= \cosh((t-s)H) - 2^{2H-1} |\sinh((s-t)/2)|^{2H} =: \vphi(s-t).  
\end{align*}
We then have 
\[ C(s,-t) = \vphi(s+ t) = \cosh((t+s)H) -2^{2H-1}  |\sinh((s+t)/2)|^{2H}\]
and 
\begin{align*}
2\vphi(x) 
&= e^{Hx} + e^{-Hx} -(e^{x/2} - e^{-x/2})^{2H} 
= e^{Hx} + e^{-Hx} - e^{Hx}(1 - e^{-x})^{2H} \\
&= e^{-Hx} + e^{Hx}\sum_{k = 1}^\infty {2H \choose k} (-1)^{k-1} e^{-kx}.
\end{align*}
In Proposition~\ref{prop:6.1} we have seen that 
this function is positive definite if and only if 
$0 < H \leq 1/2$. 
Hence $C$ is reflection positive for $0<H\leq 1/2$.

Now let $H=1/2$ so that $C(t, s)$ corresponds to the Ornstein--Uhlenbeck process. 
In this case $C(t , s)$  is invariant 
under the reflection $\theta(t) = -t$. 
As $\chi_{[0,1]}$ is cyclic in $L^2(\R_+)$ for the dilation group, 
there exists a unique unitary isometry 
$V$ on $L^2(\R_+)$ with $V(\tau_{e^t}\chi_{[0,1]}) = \tau_{e^{-t}} \chi_{[0,1]}$ 
for $t \in \R$. The latter relation is is equivalent to 
$e^{t/2} V(\chi_{[0,e^{-t}]}) = e^{-t/2} \chi_{[0,e^t]},$ resp., 
\begin{equation}
  \label{eq:or-uhl}
V(\chi_{[0,t]}) = t \chi_{[0,t^{-1}]} \quad \mbox{ for } \quad t > 0.
\end{equation}
Therefore $V$ coincides with the unitary involution $\hat\theta$ 
corresponding to the symmetry of Brownian motion under inverstion of~$t$  
(see Remark~\ref{rem:ch-trafo}(a), and also Lemma~\ref{lem:8.3} below).

Note that 
$(\theta\xi)(x) = \frac{1}{x} \xi\big(\frac{1}{x}\big)$ 
also defines an isometric involution on $L^2(\R_+^\times)$ having the same 
intertwining properties with the dilation group as $\hat\theta$,  
but this involution does not fix $\chi_{[0,1]}$.

To derive a formula for the involution $\hat\theta$, 
we recall the Sobolev space $H^1_*(\R)$. 

 \begin{defn} Let $H^1_*(\R)$ denote the Sobolev space of all absolutely continuous 
functions $F \: \R \to \R$ satisfying $F(0)= 0$ and $F' \in L^2(\R)$. Then 
\[ I \: L^2(\R_+) \to H^1_*(\R), \quad 
I(f)(t) := \int_0^t f(s)\, ds  = \la b_t^{1/2}, f \ra\] 
is a bijection. We define a real Hilbert space structure on $H^1_*(\R)$ in  such a 
way that $I$ is isometric. The inverse isometry is then given by $F \mapsto F'$.
\end{defn}

\begin{rem} (a) From the relation 
$I(f)(t) = \la f, b^{1/2}_t\ra$, it follows that 
$H^1_*(\R)$ is the real reproducing kernel Hilbert space with 
kernel $C = C^{1/2}$, i.e., the covariance kernel of Brownian motion $(B_t)_{t \in \R}$. 

(b) We observe that 
$|F(t)| \leq \|F\| |\sqrt{t}|$ for 
$F \in H^1_*(\R)$ and $t \in \R$ follows immediately 
from the Cauchy--Schwarz inequality and $\|b_t^{1/2}\|_2^2 
= C^{1/2}(t,t)= |t|$. 
\end{rem}

\begin{lem} \mlabel{lem:8.3} 
There exists a uniquely determined isometric involution 
$\hat\theta$ on $L^2(\R_+)$ satisfying 
\[ \hat\theta(\chi_{[0,t]}) = t \chi_{[0,t^{-1}]} \quad \mbox{ for } \quad t > 0.\]
It is given by 
\begin{equation}
  \label{eq:theta}
(\hat\theta \xi)(t) = \int_0^{1/t} \xi(s)\, ds - \frac{1}{t} \xi\Big(\frac{1}{t}\Big) 
\quad \mbox{ for } \quad t > 0.
\end{equation}
\end{lem}

\begin{prf} First we observe that the family  $b_t = \chi_{[0,t]}$, $t > 0$, 
is total in $L^2(\R_+)$. 
Since the family $\tilde b_t := t b_{1/t}$  satisfies
$\la \tilde b_t, \tilde b_s \ra = \la b_t, b_s \ra 
= t \wedge s$ (Remark~\ref{rem:ch-trafo}), there exists a uniquely determined isometry 
$\hat\theta$  with $\hat\theta(b_t) = \tilde b_t$ for 
$t > 0$. As $(\tilde b_t)_{t > 0}$ is also total, 
$\hat\theta$ is surjective. Now $\hat\theta(\tilde b_t) = t \frac{1}{t} b_t= b_t$ 
for $t > 0$ implies $\hat\theta^2 = \1$. 

Let $\tilde \theta$ denote the involutive isometry of $H^1_*(\R)$ specified by 
$\tilde \theta \circ I = I \circ \hat\theta$, resp., 
$\hat\theta(F') = \tilde\theta(F)'$. 
Then we obtain 
\[ (\tilde \theta I(f))(t) 
= I(\hat \theta f)(t) 
= \la \hat \theta f, \chi_{[0,t]} \ra  
= \la f, \hat\theta\chi_{[0,t]} \ra  
= t\la f,  \chi_{[0,t^{-1}]} \ra  
= t I(f)(t^{-1})\] 
and thus 
\[ (\tilde \theta F)(t) = t F(t^{-1}).\]
For $f = F'$, this leads to 
\[ (\hat\theta f)(t) = (\hat\theta F')(t) 
= (\tilde\theta F)'(t)
= F(t^{-1}) - \frac{1}{t} F'(t^{-1}) 
= \int_0^{t^{-1}} f(s)\, ds- \frac{1}{t} f(t^{-1}).\] 
This proves \eqref{eq:theta}. 
\end{prf} 

\begin{rem} (a) 
Note that \eqref{eq:theta} has a striking similarity with the formula 
for $T_J$ one finds in \cite[p.~273]{Ta88b}. 
This suggests these operators correspond to a discrete series representation 
of $\SL_2(\R)$, hence cannot be implemented in the complementary series 
representation that we consider. 

(b)
From the explicit formula for $\hat\theta$, we can also make the 
natural map $q$ from $\cE_+ := L^2([0,1]) \subeq \cE = L^2(\R_+)$ to the 
space $\hat\cE\cong \C$ more explicit. It is given by 
\[ q(f) = \int_0^1 f(x)\, dx.\] 
This follows from 
\[ \la f, \hat\theta  f \ra 
= \int_0^1 \int_0^{1/x} f(u)\, du\, \oline{f(x)}\, dx 
- \int_0^1 \oline{f(x)} \frac{1}{x} f(x^{-1})\, dx 
= \int_0^1 \int_0^1 f(u)\oline{f(x)}\, du 
= \Big|\int_0^1 f(x)\, dx\Big|^2.\] 
Here $\chi_{[0,1]} \in \cE_+$ spans the one-dimensional subspace of 
$\hat\theta$-fixed points, so that 
$q \: \cE_+ \to \hat\cE$ can be identified with the projection 
onto $\C \chi_{[0,1]}$. 
\end{rem}
 
Takenaga's formula (\cite{Ta88a})
\[ B^g_t := (ct + d) B_{g.t} - ct \cdot B_{g.\infty} - d \cdot B_{g.0}, \qquad 
g = \pmat{a & b \\ c &d} \in \SL_2(\R), \qquad t \in \R_\infty, 
g.t \not=\infty,\] 
defines for each $g \in \SL_2(\R)$ on $H^1_*(\R) \cong \cH_{C^{1/2}}$ 
a unitary operator which acts on the point evaluations 
$(b_t)_{t \in \R}$ by 
\[ U_g^{-1} b_t := b^g_t := (ct + d) b_{g.t} - ct \cdot b_{g.\infty} - d \cdot b_{g.0}, \] 
where we put 
$c b_{g.\infty} = 0$ for $c = 0$ ($g.\infty = \infty$),  
$d \cdot b_{g.0} = 0$ for $d = 0$ ($g.0 = \infty$) and 
$(ct + d) b_{g.t} =0$ for $ct + d= 0$ ($g.t = \infty$). 
On general functions $F \in H^1_*(\R)$, the operator $U_g$ acts by 
\[  (U_g F)(t) = \la b_t, U_g F \ra = \la U_g^{-1} b_t, F \ra 
= (ct+d) F(g^{-1}.t) - ct F(a/c) - d F(b/d), \qquad 
g^{-1} = \pmat{a & b \\ c &d}.\] 
Note that all summands are well defined for $c = 0$, $d = 0$, resp., $ct + d =0$ 
because $|F(t)| \leq \|F\| \cdot |t|^{1/2}$ implies 
\[ \lim_{t \to 0} t F(s/t) = 0 \quad \mbox{ for }\quad s \in \R.\] 
The relation $(b^g)^h_t = b^{gh}_t$ for $g,h \in \SL_2(\R)$ now leads to 
\[ U_{gh}^{-1} b_t = b_t^{gh} = U_h^{-1} b_t^g = U_h^{-1} U_g^{-1} b_t 
\quad \mbox{ for } \quad g,h \in \SL_2(\R), t \in \R,\] 
and hence to $U_g U_h = U_{gh}$. We thus obtain on $H^1_*(\R)$ 
a continuous unitary representation of $\SL_2(\R)$. Note that 
$U_{-\1} = -\1$, so that this representation does NOT factor through 
a representation of $\PSL_2(\R)$.
\begin{footnote}
  {The most economical way to verify the assertion that the operators 
$U_g$ are unitary is to do that for $g.t = at + b$ and 
$\sigma.t = -t^{-1}$ and then to verify that 
$(b^g)^h_t = b^{gh}_t$ holds for all $g \in \SL_2(\R)$ and 
$h.t = at + b$ or $h.t = - t^{-1}$. As $\SL_2(\R)$ is generated by 
elements of this form, it follows that $U$ defines a unitary representation 
on $H^1_*(\R) \cong \cH_{C^{1/2}}$.}
\end{footnote}

With the aforementioned conventions concerning expressions of the form 
$t F(s/t) =0$ for $t = 0$, the representation 
is given by 
\begin{align} 
  \label{eq:tak}
 (U_g F)(t) 
&= (ct+d) F(g^{-1}.t) - ct F(a/c) - d F(b/d) \notag \\
&=  (ct+d) F(g^{-1}.t) - ct \cdot F(g^{-1}.\infty) - d \cdot F(g^{-1}.0).
\end{align} 

If $g^{-1}.t = \alpha t + \beta$ with $\alpha > 0$ is affine, then 
$g^{-1} = \pmat{ \sqrt{\alpha} & \sqrt{\alpha}\beta \\ 
0 & \sqrt{\alpha}^{-1}}$ and we get 
\[  (U_g F)(t) = \alpha^{-1/2} (F(\alpha t + \beta) - F(\beta)).\] 
For $J := \pmat{0 & -1 \\ 1 & 0}$, we have $J^{-1} = -J$, so that 
\[  (U_J F)(t) 
= -t (F(-t^{-1}) - F(0)) = -t F(-t^{-1}).\] 

We also note that \eqref{eq:tak} leads to 
\begin{align*}
  (U_g F)'(t) 
&= c (g_* F)(t) + (ct+d) (g_* F')(t)\frac{1}{(c t + d)^2} - c F(a/c)\\
&= c\big(F(g^{-1}.t) - F(g^{-1}.\infty)\big)+ (ct+d)^{-1} (g_* F')(t).
\end{align*}
If $g^{-1}.t= \alpha t+ \beta$ is affine, then
\[  (U_g F')(t) = \alpha^{1/2} F'(\alpha t + \beta) \] 
yields the usual action of $\Aff(\R)$ on $L^2(\R)$. 

\begin{rem}
Since we want to express this in terms of the derivatives, we observe that, 
formally,  we expect something like
\begin{align*}
 F(g^{-1}.t) - F(g^{-1}.\infty)
&= \int_{g^{-1}.\infty}^{g^{-1}.t} F'(x)\, dx 
= \int_{\infty}^{t} F'(g^{-1}.x)\, (g^{-1})'(x)\, dx \\
&= -\int_t^{\infty} \frac{F'(g^{-1}.x)}{(c x + d)^2}\, dx
= \int_{-\infty}^t \frac{F'(g^{-1}.x)}{(c x + d)^2}\, dx.  
\end{align*}
\end{rem}

In particular, we have 
\[  (U_J F)'(t) 
= -F(-t^{-1}) - t \cdot t^{-2} F'(-t^{-1})
= - F(-t^{-1}) - t^{-1} F'(-t^{-1})
= -\int_0^{-t^{-1}} F'(x)\, dx -   t^{-1} F'(-t^{-1}).\] 
For $\xi = F'$, this reads 
\[ (U_J \xi)(t) =  - t^{-1} \xi(-t^{-1}) - \int_0^{-t^{-1}} \xi(x)\, dx
\quad \mbox{ for } \quad t\not=0.\] 
This formula describes the unique unitary involution on $L^2(\R)$ mapping 
$b_t$ to $b_{-t^{-1}}$ 
(cf.\ Lemma~\ref{lem:8.3}). 
 
\appendix 

\section{Existence of affine isometries} 
\mlabel{sec:a.2}

For a map $\gamma \: X \to \cH$ into a Hilbert space, 
the closed subspace $\cH_\gamma$ generated by all differences 
$\gamma(x)-\gamma(y)$, $x,y \in X$, is called the {\it chordal 
space of $\gamma$} (cf.\ \cite{Fu05}). 
The following lemma is an abstraction of \cite[Satz~1,3]{Ko40b}. 

\begin{lem} \mlabel{lem:ex-isom} Let $X$ be a non-empty set, 
$\cH$ be a real or complex Hilbert space and 
$\gamma\: X \to \cH$ 
and $\gamma'\: X \to \cH'$ 
be maps with $\cH_\gamma = \cH$ and $\cH_{\gamma'} = \cH'$. 
For $x_0 \in X$, consider the kernel 
\[ K^{x_0}_\gamma(x,y) := \la \gamma(x) - \gamma(x_0), \gamma(y) - \gamma(x_0)\ra 
\quad \mbox{ on } \quad X \times X.\] 
Then the following are equivalent: 
\begin{itemize}
\item[\rm(i)] There exists an affine isometry $V \:  \cH \to \cH'$ with 
$V\circ \gamma = \gamma'$. 
\item[\rm(ii)] $K^{x_0}_\gamma = K^{x_0}_{\gamma'}$ 
for every $x_0 \in X$. 
\item[\rm(iii)] $K^{x_0}_\gamma = K^{x_0}_{\gamma'}$ 
for some $x_0 \in X$.   
\end{itemize}
If $\cH$ and $\cH'$ are real, then these conditions are equivalent to 
\begin{itemize}
\item[\rm(iv)] $\|\gamma(x)-\gamma(y)\|^2 = \|\gamma'(x)-\gamma'(y)\|^2$ for 
$x,y \in X$. 
\end{itemize}
If {\rm(i)-(iii)} are satisfied, then the affine 
isometry $V$ in {\rm(i)} is uniquely determined by the relation $V \circ \gamma = \gamma'$. 
\end{lem}

\begin{prf} (ii) $\Rarrow$ (iii) is trivial.

(iii) $\Rarrow$ (i): From \cite[Ch.~I]{Ne00} it follows that there exists a
 unique unitary operator $U \: \cH \to \cH$ with 
\[ U(\gamma(x)- \gamma(x_0)) = \gamma'(x) - \gamma'(x_0) \quad \mbox{ for all } \quad 
x \in X.\] 
Then we put $V\xi := U\xi - U(\gamma(x_0)) + \gamma'(x_0)$.

(i) $\Rarrow$ (ii): If $V\xi = U\xi + b$ is an affine isometry with 
$V\circ \gamma = \gamma'$, then 
\begin{align*}
 K^{x_0}_{\gamma'}(x,y) 
&=\la \gamma'(x) - \gamma'(x_0), \gamma'(y) - \gamma'(x_0)\ra 
=\la U\gamma(x) - U\gamma(x_0), U\gamma(y) - U\gamma(x_0)\ra \\ 
&=\la \gamma(x) - \gamma(x_0), \gamma(y) - \gamma(x_0)\ra = K^{x_0}_\gamma(x,y).
\end{align*}

(iv) $\Leftrightarrow$ (iii): The kernel 
$D_\gamma(x,y) := \|\gamma(x)-\gamma(y)\|^2$ 
satisfies 
\begin{align*}
 D_\gamma(x,y) 
&= \|\gamma(x)-\gamma(x_0) + \gamma(x_0) - \gamma(y)\|^2  \\ 
&= \|\gamma(x)-\gamma(x_0)\|^2  + \|\gamma(x_0) - \gamma(y)\|^2  
+ 2 \Re K_\gamma^{x_0}(x,y) \\ 
&= K_\gamma^{x_0}(x,x) +K_\gamma^{x_0}(y,y) +2 \Re K_\gamma^{x_0}(x,y)
\end{align*}
and, conversely,
\[ \Re K_\gamma^{x_0}(x,y) 
= \frac{1}{2}\big( D_\gamma(x,y) - D_\gamma(x,x_0) - D_\gamma(y,x_0)\big).\] 
Therefore (iv) is equivalent to 
$\Re K_\gamma^{x_0} = \Re K_{\gamma'}^{x_0}$. If $\cH$ is real, 
this is equivalent to (iii). 
\end{prf}

\section{Stochastic processes} 
\mlabel{sec:a.1}

\begin{defn} \mlabel{def:st-pro} 
Let $(Q,\Sigma,\mu)$ be a probability space and $(B,\fB)$ be a measurable space. 
A {\it stochastic process} with state space $(B,\fB)$ 
is a family $(X_t)_{t \in T}$ of measurable 
functions $X_t \: Q \to B$, where $T$ is a set. 

(a) We call the stochastic process $(X_t)_{t \in T}$ {\it full} if, up to sets 
of measure $0$, $\Sigma$ is the smallest $\sigma$-algebra for which 
all functions $X_t$ are measurable. 

(b) For $B = \R$ or $\C$, 
we say that $(X_t)_{t \in T}$ is {\it square integrable} if every $X_t$ 
is square integrable. Then the {\it covariance kernel} 
\[ C(s,t) := \bE(\oline{X_s} X_t) \]
on $T$ is positive definite. 
If $C(t,t) = \E(|X_t|^2) > 0$ for every $t \in  T$, 
then 
$\tilde X_t := X_t/\sqrt{\E(|X_t|^2)}$ is called the {\it associated normalized 
process}. Its covariance kernel is 
\[ \tilde C(s,t) = \frac{C(s,t)}{\sqrt{C(s,s)C(t,t)}} 
\quad \mbox{ for } \quad s,t \in T.\]

(c) On the {\it product space} $B^T$ of all maps $T \to B$, 
there exists a unique probability measure $\nu$ 
with the property that, for $t_1, \ldots, t_n \in T$, the image of $\nu$ under the evaluation map 
$\ev_{t_1,\ldots, t_n} \: B^T \to B^n$ is the image of $\mu$ under the map 
$(X_{t_1}, \ldots, X_{t_n})$. We call $\nu$ the {\it distribution of the process 
$(X_t)_{t \in T}$} (\cite[Thm.~1.5]{Hid80}). 
\end{defn}

\begin{defn} \mlabel{def:b.5}
Let $(X_t)_{t \in T}$ be a centered $\K$-valued stochastic process 
and $\sigma \: G \times T \to T$ be a group acting on $T$. 

(a) The process $(X_t)_{t \in T}$ is called {\it stationary} 
if, for every $g \in G$, the process 
$(X_{g.t})_{t \in T}$ has the same distribution. Then we obtain a 
measure preserving $G$-action on the underlying path space $\K^T$ by 
$(g.\omega)(t) := \omega(g^{-1}.t)$, resp., $g.X_t = X_{g.t}$. 

(b) The process $(X_t)_{t \in T}$ is said to have {\it stationary
increments} 
if, for $t_0, t_1, \ldots, t_n, t \in T$, the random vectors 
\[ (X_{t_1} - X_{t_0}, \ldots, X_{t_n}- X_{t_0}) \quad \mbox{ and } \quad 
(X_{g.t_1} - X_{g.t_0}, \ldots, X_{g.t_n}- X_{g.t_0}) \] 
have the same distribution (cf.\ \cite{Ko40a}). 
\end{defn}

\begin{defn} \mlabel{def:widesensestat} (\cite[Def.~2.8.1]{Sa13}) 
  A square integrable 
process $(Z_t)_{t \geq 0}$ is said to be {\it wide sense stationary} if the 
function $t \mapsto \E(Z_t)$ is constant and there exists a function $C\: \R\to \C$ such
that
$C(s,t) = \E(\oline{Z_s}Z_t) = C(s-t)$.
\end{defn}

\subsection{Processes with stationary increments} 
\mlabel{subsec:statinc}

\begin{prop} {\rm(The flow of a process with stationary increments)}
Let $(X_t)_{t \in T}$ be a $\K$-valued stochastic process 
and $\sigma \: G \times T \to T, (g,t) \mapsto g.t$ be a $G$-action on~$T$. 
Then the following are equivalent: 
\begin{itemize}
\item[\rm(i)] $(X_t)_{t \in T}$ has stationary increments. 
\item[\rm(ii)] For every $t_0 \in T$, 
  \begin{equation}
    \label{eq:pathspace-act}
(g.\omega)(t) := \omega(g^{-1}.t) +  \omega(t_0) - \omega(g^{-1}.t_0) 
  \end{equation}
defines a measure preserving flow on the path space $\K^T$ satisfying 
\begin{equation}
  \label{eq:trafo-g}
g.X_t = X_{g.t} + X_{t_0} - X_{g.t_0}. 
\end{equation}
\end{itemize}
\end{prop}

\begin{prf} (i) $\Rarrow$ (ii): For each $g \in G$, we consider the map 
\begin{equation}
  \label{eq:y1}
\oline\sigma_g \: \K^T \to \K^T, \quad \sigma_g(\omega)(t) 
:= \omega(g^{-1}.t) +  \omega(t_0) - \omega(g^{-1}.t_0). 
\end{equation}
Then 
\begin{equation}
  \label{eq:y2}
((\oline\sigma_g)_*X_t)(\omega) = X_t(\oline\sigma_{g^{-1}}\omega) 
= \omega(g.t) + \omega(t_0) - \omega(g.t_0),\
\end{equation}
i.e., 
\[ (\sigma_g)_*X_t = X_{g.t} + X_{t_0} - X_{g.t_0}, \quad \mbox{ resp., } \quad 
X_{g.t} = (\sigma_g)_* X_t + X_{g.t_0}- X_{t_0}.\] 
Since, for every finite subset $F \subeq T$, the random 
vector $(X_{g.t} - X_{g.t_0})_{t \in F}$ has the same distribution as 
$(X_t - X_{t_0})_{t \in F}$, the flow on $\K^T$ defined by $\sigma$ is measure preserving. 

(ii) $\Rarrow$ (i): If there exists a measure preserving $G$-action 
on $\K^T$ satisfying \eqref{eq:trafo-g},
then the distribution of 
$(X_{g.t} + X_{t_0} - X_{g.t_0})_{t \in T}$ is the same as the distribution of 
$(X_t)_{t \in T}$. Subtracting $X_{t_0}$, it follows that the distribution of 
$(X_{g_t} - X_{g.t_0})_{t \in T}$ is the same as the distribution of 
$(X_t - X_{t_0})_{t \in T}$, i.e., that $(X_t)_{t \in T}$ has stationary increments. 
\end{prf}

\begin{rem} \mlabel{rem:1.13} 
(a) If the $\K$-valued 
process $(X_t)_{t \in T}$ on $(Q,\Sigma, \mu)$ is square integrable, then 
$(X_t)_{t\in T}$ generates a closed linear subspace $\cH_1 \subeq L^2(Q,\mu)$. 
The existence of a unitary representation 
$(U_g)_{g \in G}$ on $\cH_1$ with $U_g X_t = X_{g.t}$ for $g \in G, t \in T$, 
is equivalent to the invariance of the covariance kernel 
\[ C(s,t) := \bE(\oline{X_s}X_t ) = \la X_s, X_t \ra \] 
(cf.\ \cite[Ch.~I]{Ne00}). 
This condition is in particular satisfied if the process is stationary. 

(b) For a square integrable process, 
it likewise follows that the existence of an action of $G$ by affine isometries 
$(\alpha_g)_{g \in G}$ on  the closed affine subspace $\cA\subeq L^2(Q,\mu)$ 
generated by $(X_t)_{t \in T}$ satisfying 
\[ X_{g.t} = \alpha_g X_t  \quad \mbox{ for } \quad g \in G, t \in T\] 
is equivalent to the independence from $g \in G$ of the kernel 
\[ Q^g(t,s) := \E((X_{g.t}-X_{g.t_0})(\oline{X_{g.s}-X_{g.t_0}})) \quad 
\mbox{ for } \quad s,t \in T, \] 
for some $t_0\in T$ (and hence for all $t_0 \in T$)) 
(Lemma~\ref{lem:ex-isom}).
For a real-valued process ($\K = \R$), this condition is equivalent to the 
$G$-invariance of the kernel 
\[ D(t,s) := \E((X_t - X_s)^2)\quad \mbox{ for } \quad t,s \in T \] 
on $T \times T$ (Lemma~\ref{lem:ex-isom}).
\end{rem}

\begin{lem} \mlabel{lem:b.6} Let $(\alpha_t)_{t \in \R}$ be a continuous 
isometric affine $\R$-action of the form 
\[ \alpha_t\xi = U_t \xi + \beta_t\quad \mbox{ for } \quad t \in \R,\xi \in \cH\] 
on the real or complex Hilbert space $\cH$.  
If $\beta_\R$ is total in $\cH$, 
then the unitary representation $(U_t)_{t \in \R}$ is cyclic. 
\end{lem}

\begin{prf} {\bf First proof:} We write $\cH = \cH_0 \oplus \cH_1$, where 
$\cH_0 = \cH^U$ is the closed subspace of $U$-fixed vectors and 
$\cH_1 := \cH_0^\bot$. Accordingly, we write 
$\beta = \beta_0 + \beta_1$. 
Then $\beta_0 \: \R \to \cH_0$ is a continuous homomorphism, hence of the 
form $\beta_0(t) = t v_0$ for some $v_0 \in \cH_0$. We conclude that 
$\dim \cH_0 \leq 1$, so that it suffices to show that the representation 
on $\cH_1$ is cyclic. 
We may therefore assume from now on that $\cH^U = \{0\}$. 

{\bf Step 1:} First we assume that $\Spec(U)$ is compact and does not contain~$0$. 
Then there exists an $\eps > 0$ such that the operators $U_t -\1$ are invertible 
for $|t| < \eps$. For $|t|,|s| < \eps$, we then have 
\[ (U_t - \1) \beta_s  = \beta_{t+s} - \beta_s - \beta_t = (U_s - \1) \beta_t,\] 
so that 
\[ v := (U_t - \1)^{-1} \beta_t\quad \mbox{ for } \quad |t| < \eps \] 
is independent of $t$. Now the relation 
$\beta_t = U_t v - v$ holds for $|t| < \eps$, but since $\beta$ is a continuous 
cocycle, it follows for all $t \in \R$. Clearly, $v \in \cH$ is a $U$-cyclic vector. 

{\bf Step 2:} Now we consider the general case where $\cH$ is complex. 
We write $\R^\times = \bigcup_{n \in \N} C_n$, where $C_n$ 
is relatively compact with $0 \not\in \oline{C_n}$. 
If $P$ is the spectral measure of $U$, we accordingly obtain a $U$-invariant 
decomposition $\cH = \hat\oplus_{n \in \N} P(C_n)\cH$ into subspace on which 
$U$ has compact spectrum not containing~$0$. Now our assumption implies that 
every $\cH_n$ is generated by the values of the $\cH_n$-component of $\beta$. 
Step 1 now implies that each $\cH_n$ is cyclic, and since representations on the 
subspaces $\cH_n$ are mutually disjoint,  the representation 
on $\cH$ is cyclic. 

{\bf Step 3:} Finally, we consider the general case where $\cH$ is real. 
Then we may choose the sets $C_n \subeq \R$ such that they are symmetric, i.e., 
 $C_n = - C_n$. Then the corresponding spectral subspaces 
of $\cH_\C$ are invariant under complex conjugation and we can proceed as 
in Step 2. 

{\bf Alternative proof:} A more direct argument can be derived from the 
work of P.~Masani (\cite{Mas72}; see also \cite{Fu05}). 
For the element 
\[ \xi := \int_0^\infty e^{-t} \beta_t\, dt \] 
one shows that the {\it shift operators} 
\[ T(a,b) := U_b - U_a -\int_a^b U_t\, dt  = - T(b,a) \] 
satisfy $\beta_t = T(t,0) \xi$. Here the main point is to verify first the 
{\it switching property} (\cite[Lemma~2.18]{Mas72}) 
\[ T(a,b)(\beta_c - \beta_d) = T(c,d)(\beta_a - \beta_b)
\quad \mbox{ for } \quad a,b,c,d \in \R,\] 
and that $\int_0^\infty e^{-t} T_U(t,0)\, ds = \1$ 
(\cite[Thm.~A.2]{Mas72}). 
Then the assertion follows from 
\[ T(t,0)\xi 
= \int_0^\infty e^{-s} T(t,0)\beta_s\, ds 
= \int_0^\infty e^{-s} T(s,0)\beta_t\, ds = \beta_t \] 
(\cite[Thm.~2.19]{Mas72}). 
\end{prf}

\begin{prop} \mlabel{prop:normform} {\rm(Normal form of cocycles)}
Let $(U,\cH)$ be a continuous unitary one-parameter group 
and $\beta \: \R \to \cH$ be a continuous cocycle. 
Then there exists a Borel measure $\sigma$ on $\R$ such that the 
triple $(U,\beta,\cH)$ is unitarily equivalent to the triple 
$(\tilde U, \tilde\beta, L^2(\R,\sigma))$ with 
\[ (\tilde U_t f)(x) = e^{itx} f(x) \quad \mbox{ and } \quad 
\beta_t(x) =\begin{cases}
\frac{e^{itx}-1}{ix} & \text{for }  x\not=0 \\   
t & \text{for }  x=0.\end{cases}
\] 
\end{prop}

\begin{prf} In view of Lemma~\ref{lem:b.6}, we may assume that 
the representation $(U,\cH)$ is cyclic. 

{\bf Step 1:} First we assume that $\cH^U = \{0\}$. 
According to Bochner's Theorem, any cyclic unitary one-parameter group 
$(U,\cH)$ with $\cH^U = \{0\}$ 
is equivalent to the multiplication representation on some 
space $L^2(\R^\times,\mu)$ by $(U_t f)(x) = e^{itx} f(x)$. For this representation 
it is easy to determine the cocycles. They are of the form 
\[ \beta_t(x) = (e^{itx}-1)u(x), \] 
where $u \: \R \to \C$ is a measurable function with the property that, for 
every $t \in  \R$, the function $(e^{itx} -1)u$ is square integrable. 
Replacing $\mu$ by the measure 
\[ d\sigma(x) = x^2 |u(x)|^2\, d\mu(x),\] 
we may assume that $u(x) = \frac{1}{ix}$, which leads to 
$\beta_t(x) = \frac{e^{itx}-1}{ix}$. 

{\bf Step 2:} If $\cH^U = \cH$, then $\beta \: \R \to \cH$ is a continuous 
homomorphism, hence of the form $\beta_t= t v$ for some $v \in \cH$. 
The cyclicity assumption implies that $\cH = \C v 
\cong L^2(\R, \sigma)$ for the measure $\sigma = \|v\|^2 \delta_0$. 
Here the vector $v$ corresponds to the constant function $1$, 
so that $\beta_t = t v = t$. 

The assertion now follows by applying Steps 1 and 2 to the summands of 
the decomposition $\cH = \cH^U \oplus (\cH^U)^\bot$. 
\end{prf}

The following theorem is basically the L\'evy--Khintchine Theorem 
for the group $G = \R$ (cf.\ \cite[Thm~5.5.1]{Luk70}, 
\cite[Thm.~32]{Le65}, and \cite{Ko40b} for a different form). 

\begin{prop}
  \mlabel{prop:a.4b} 
Let $(X_t)_{t \in \R}$ be a complex-valued zero mean Gaussian process on 
$(Q,\Sigma,\mu)$ with $X_0 = 0$ and stationary 
quadratic increments. Then there exists a uniquely determined 
Borel measure $\sigma$ on $\R$ such that 
\begin{equation}
  \label{eq:k-form}
C_\sigma(s,t) = \bE(X_s^*X_t ) = \int_{\R} \oline{e_s(u)}e_t(u)\, d\sigma(u)
\quad \mbox{ for } \quad 
e_t(u) = \frac{e^{itu}-1}{iu} = \int_0^t e^{i\tau u}\, d\tau.
\end{equation}
A measure $\sigma$ on $\R$ arises for such a process if and only if
 
\begin{equation}
  \label{eq:fincond}
\int_\R \frac{d\sigma(u)}{1 + u^2} < \infty.
\end{equation}
The function 
\[ r(t) 
:= \int_\R \Big(1 - e^{itu} + \frac{itu}{1 + u^2}\Big)\frac{d\sigma(u)}{u^2} \] 
is negative definite and satisfies 
\begin{equation}
  \label{eq:r-rel}
r(t) + \oline{r(s)} - r(t-s) = C_\sigma(s,t).
\end{equation} 
All other negative definite continuous functions satisfying 
\eqref{eq:r-rel} are of the 
form $\tilde r(t) = r(t) + i t\mu$ for some $\mu \in \R$. 
\end{prop}

The measure $\sigma$ is called the {\it spectral measure} 
of the process $(X_t)_{t \in \R}$. 

\begin{prf} Let $\cA \subeq L^2(Q,\Sigma,\mu)$ be the closed affine subspace 
generated by $(X_t)_{t \in \R}$. As $X_0 = 0$, this is actually 
a linear subspace. Now Lemma~\ref{lem:ex-isom} 
implies the existence of an affine isometric action 
$(\alpha_t)_{t \in \R}$ of $\R$ on $\cA$ satisfying $\alpha_t X_s = X_{s+t}$. 
In particular, $\beta_t := X_t$ is a corresponding cocycle. 
Now the existence of $\sigma$ follows from Proposition~\ref{prop:normform}. 

Now we show that \eqref{eq:fincond} 
is equivalent to the square integrability of all $(e_t)_{t \not=0}$ 
(Definition~\ref{def:st-pro}) and the continuity of the 
function $t \mapsto C_\sigma(t,t) = \|e_t\|_2^2$. 

From 
\[ |e_t(u)|^2 
= \Big|\frac{\cos(tu) - 1}{u}\Big|^2 + \Big|\frac{\sin(tu)}{u}\Big|^2
= \frac{1 + \cos^2(tu) + \sin^2(tu) - 2 \cos(tu)}{u^2}
=2 \frac{1- \cos(tu)}{u^2}\]
it follows that the square integrability of all $e_t$ with respect to 
$\sigma$ is equivalent to 
\[ f(t) := \int_\R \frac{1- \cos(tu)}{u^2}\, d\sigma(u) < \infty 
\quad \mbox{ for all } \quad  t\in \R.\] 
If $r > 0$ and $t$ is sufficiently small, then the integrand 
has a positive infimum on the interval $[-r,r]$. 
Therefore the finiteness of all $f(t)$ implies that 
all compact subsets of $\R$ have finite $\sigma$-measure. 
Since the function $f(t) = \frac{1}{2} C_\sigma(t,t)$ is continuous, 
for every $\eps > 0$, we have
\[ \infty > \int_0^\eps f(t)\, dt 
= \int_\R \int_0^\eps (1 - \cos(tu))\, dt\, \frac{d\sigma(u)}{u^2} 
= \int_\R \Big(\eps - \frac{\sin \eps u}{u}\Big) \, \frac{d\sigma(u)}{u^2}.\] 
As the function $u \mapsto 1 - \frac{\sin (\eps u)}{\eps u}$ 
has a positive infimum on $[1,\infty)$, it follows that 
$\int_{|u| \geq 1} \frac{d\sigma(u)}{u^2} < \infty$.
This implies that $\int_\R  \frac{d\sigma(u)}{1 + u^2} < \infty$.

Suppose, conversely, $\int_\R  \frac{d\sigma(u)}{1 + u^2} < \infty$ 
(cf.\ \cite[Lemma~5.5.1]{Luk70}). 
We claim that we obtain a continuous negative definite function 
\begin{equation}
  \label{eq:l-k}
r(t) 
:= \int_\R \Big(1 - e^{itu} + \frac{itu}{1 + u^2}\Big)\frac{d\sigma(u)}{u^2} 
= \int_\R \Big(\frac{1 - e^{itu}}{u} + \frac{it}{1 + u^2}\Big)\frac{d\sigma(u)}{u}. 
\end{equation}
We first show that the integrals exist. 
To this end, we observe that 
\begin{align*}
& \Big(1 - e^{itu} + \frac{itu}{1 + u^2}\Big)\frac{1 + u^2}{u^2}  
= 1 - e^{itu} + \frac{1 + itu - e^{itu}}{u^2}. 
\end{align*}
Since all three summands are bounded, the existence of the 
integral \eqref{eq:l-k} 
defining $r(t)$ follows. The first two summands are bounded independently 
of $t$, and the third summand can also be written as 
\[ \frac{1 + itu - e^{itu}}{u^2}  = h(tu) t^2,\] 
where the function $h \: \R \to \C$ is bounded. We conclude that all summands 
are locally uniformly bounded in $t$. Therefore the continuity of the function 
$r$ follows  from Lebesgue's Dominated Convergence Theorem. 
Moreover, $r$ is negative 
definite because the functions $t \mapsto 1 - e^{itu}$ and $t \mapsto it$ are. 

We further have the relation 
\begin{equation}
  \label{eq:k-r}
r(t) + \oline{r(s)} - r(t-s) = 
C_\sigma(s,t) = \bE(X_s^* X_t) = \int_{\R} \oline{e_s(u)} e_t(u)\, d\sigma(u), 
\end{equation}
showing that $C_\sigma$ is the positive definite kernel associated to the 
continuous negative definite function~$r$, hence in particular continuous.  
\end{prf}

\section{Second quantization and Gaussian processes} 
\mlabel{app:2}

\begin{defn} (\cite[Def.~1.6]{Hid80})  Let $T$ be a set 
and $\K = \R$ or $\C$. 
A $\K$-valued stochastic process $(X_t)_{t \in T}$ 
is said to be {\it Gaussian} if, for all finite subsets 
$F \subeq T$, the corresponding distribution of the random vector 
$X_F= (X_t)_{t \in F}$ with values in $\K^F$ is Gaussian. 
\end{defn}

\begin{defn} \mlabel{def:1.1} Let $\cH$ be a $\K$-Hilbert space. A {\it Gaussian 
random process indexed by $\cH$} 
is a random process $(\phi(v))_{v \in \cH}$ on a probability space $(Q,\Sigma, P)$ 
indexed by $\cH$ such that 
\begin{description}
\item[\rm(GP1)] $(\phi(v))_{v \in \cH}$ is full, i.e., the random variables 
$\phi(v)$ generate the 
$\sigma$-algebra $\Sigma$ modulo zero sets. 
\item[\rm(GP2)] Each $\phi(v)$ is a Gaussian random variable of mean zero. 
\item[\rm(GP3)] $\E(\phi(v)\oline{\phi(w)}) = \la v, w\ra_\cH$ 
for $v,w \in \cH$. 
\end{description}
\end{defn}

\begin{rem} \mlabel{rem:2.1} If $T$ is a set,  $\gamma \: T \to \cH$ a map and 
$(\phi(v))_{v \in \cH}$ is a Gaussian process indexed by $\cH$, then 
$(\phi(\gamma(t)))_{t \in T}$ is a Gaussian process indexed by $T$ with 
zero means and covariance kernel 
\[ C(s,t) = \E\big(\phi(\gamma(t))\oline{\phi(\gamma(s))}\big) 
= \la \gamma(t), \gamma(s)\ra.\] 
For any function $m \: T \to \R, t \mapsto m_t$, we obtain a Gaussian 
process $(X_t)_{t \in T}$ with mean vector $(m_t)_{t \in T}$ by 
\[ X_t := \phi(\gamma(t)) + m_t.\] 
If $\gamma(T)$ is total in $\cH$, then the corresponding Gaussian process is full. 

Conversely, every Gaussian process $(X_t)_{t \in T}$ with mean vector 
$(m_t)_{t \in T}$ is of this form. Here we may choose 
$\cH$ as the subspace of $L^2(Q,\Sigma, \mu)$ generated by 
the $X_t - m_t$ (\cite[Thm.~1.10]{Hid80}).
\end{rem}

\begin{defn} (Second quantization; \cite{Si74}) \mlabel{def:3.5}
For a real Hilbert space $\cH$, we write 
$\cH^*$ for its algebraic dual, i.e., the set of all linear functionals 
$\cH\to \R$, continuous or not. Let $\Gamma(\cH) := L^2(\cH^*, \gamma,\C)$ 
denote the canonical Gaussian measure space on $\cH^*$. 
This measure is defined on the smallest $\sigma$-algebra $\Sigma = \Sigma_{\cH^*}$ 
for which all evaluations $\phi(v)(\alpha) := \alpha(v)$, $v \in \cH$, 
are measurable. It is determined uniquely by 
\[ \E(e^{i\phi(v)}) = e^{-\|v\|^2/2} \quad \mbox{ for } \quad v \in \cH.\] 
Considering the $\phi(v)$ as random variables, 
we thus obtain the canonical centered 
Gaussian process $(\phi(v))_{v \in \cH}$ over~$\cH$. 
It satisfies 
\[ \E(\phi(v)) = 0 \quad \mbox{ and } \quad 
\E(\phi(v)\phi(w)) = \la v,w \ra \quad \mbox{ for } \quad v,w \in \cH.\] 
\end{defn} 

\begin{rem} \mlabel{rem:unirep-mot} 
(The unitary representation of $\Mot(\cH)$ on $\Gamma(\cH)$) 
The group $\Mot(\cH)\cong \cH \rtimes \OO(\cH)$ of bijective isometries of $\cH$ 
has a natural unitary representation on $\Gamma(\cH)$ given by 
\[ U_{(b,g)}F = e^{i\phi(b)} g_*F,\quad \mbox{ where } \quad 
(g_*F)(\alpha) = F(g^{-1}.\alpha) = F(\alpha \circ g).\] 
In particular, the map 
\[ \cH \to \Gamma(\cH), \quad x \mapsto U_{(x,\1)}1 = e^{i\phi(x)} \] 
is $\Mot(\cH)$-equivariant with total range. 
The canonical Gaussian process over the real Hilbert space 
$\cH$ satisfies 
\begin{equation}
  \label{eq:q-ker}
Q(v,w) 
:= \E(\oline{e^{i\phi(v)}}e^{i\phi(w)}) 
= \E(e^{i\phi(w-v)}) 
= e^{-\frac{\|v-w\|^2}{2}}. 
\end{equation} 
\end{rem}

\begin{rem} 
The canonical Gaussian measure $\gamma$ on the algebraic dual 
$\cH_{1/2}^* = L^2(\R)^*$ is called {\it white noise measure}. 
The space of smooth vectors of the unitary 
representation $(U^{1/2}, L^2(\R))$ of $\GL_2(\R)$ (see \cite[Ch.~7]{NO18} for this concept) can be naturally identified 
with the space $C^\infty(\bS^1)$, considered as a subspace of $\cH_{1/2}$. 
It coincides with the space $D_0$ in Hida's book \cite[p.~304]{Hid80}. 
\end{rem}

\section{The Hilbert spaces $\cH_H$, $0 < H < 1$} 
\mlabel{app:d} 

\subsection{The scalar product on $\cH_H$} 
In this section we give a short discussion about the complementary series
representation in the one dimensional case. For detailed
discussion see \cite[pp. 28]{JOl98} and \cite[Sect. 9]{JO00}.

 For $\shalf < H < 1$ and $\xi, \eta \in \cS(\R)$, we have 
  \begin{align*}
& (2H-1) \int_{\R}\oline{\xi(x)}|x-y|^{2H-2}\, dx \\
&= (2H-1) \Big(
\int_{-\infty}^y \oline{\xi(x)}(y-x)^{2H-2}\, dx 
+ \int_y^\infty \oline{\xi(x)}(x-y)^{2H-2}\, dx\Big) \\ 
&= \int_{-\infty}^y \oline{\xi'(x)}(y-x)^{2H-1}\, dx 
- \Big[\oline{\xi(x)}(y-x)^{2H-1}\Big|_{-\infty}^y \\
&\ \ -  \int_y^\infty \oline{\xi'(x)}(x-y)^{2H-1}\, dx
+ \Big[\oline{\xi(x)}(x-y)^{2H-1}\Big|_y^\infty \\ 
&= \int_\R \oline{\xi'(x)}\frac{\sgn(y-x)}{|x-y|^{1-2H}}\, dx\, dy
  \end{align*}
We accordingly obtain 
\begin{equation}
  \label{eq:diff-form}
 \la \xi,\eta\ra_H 
= H \int_\R \int_\R \oline{\xi'(x)}\eta(y) \frac{\sgn(y-x)}{|x-y|^{1-2H}}\, dx\, dy 
\end{equation}
and thus 
\begin{align*}
\lim_{H \to \shalf}  \la \xi,\eta\ra_H 
&= \frac{1}{2}\int_\R \int_\R \oline{\xi'(x)}\eta(y)\sgn(y-x)\, dx\, dy  \\
&= \frac{1}{2}\int_\R \eta(y)\Big[
\int_{-\infty}^y \oline{\xi'(x)}\, dx - \int_y^\infty \oline{\xi'(x)}\, dx\Big] 
= \int_\R \eta(y)\oline{\xi(y)}\, dy = \la \xi,\eta\ra_{L^2(\R)}. 
\end{align*}
It therefore makes sense to put $\cH_{1/2}:= L^2(\R)$, 
so that we have Hilbert spaces $\cH_H$ for $\shalf \leq H < 1$. 

In the form \eqref{eq:diff-form}, the scalar product 
$\la\cdot,\cdot\ra_H$ is defined by a distribution kernel which 
is locally integrable for any $H > 0$. We shall use this observation to 
define Hilbert spaces $\cH_H$ for $0 < H < 1$. To find a more symmetric 
form of the scalar product, we calculate 
\begin{align*}
\int_\R \eta(y)\sgn(y-x)|x-y|^{2H-1}\, dy  
&= - \int_{-\infty}^x \eta(y)(x-y)^{2H-1}\, dy 
+ \int_x^{\infty} \eta(y)(y-x)^{2H-1}\, dy \\ 
&= \frac{1}{2H} \Big[ \eta(y)(x-y)^{2H} \Big|_{-\infty}^x 
- \frac{1}{2H}\int_{-\infty}^x \eta'(y)(x-y)^{2H}\, dy \\
&\ \ + \frac{1}{2H} \Big[ \eta(y)(y-x)^{2H} \Big|_x^{\infty} 
- \frac{1}{2H}\int_x^{\infty} \eta'(y)(y-x)^{2H}\, dy \\
&= - \frac{1}{2H}\int_\R \eta'(y)|x-y|^{2H}\, dy.
\end{align*}
We thus obtain from \eqref{eq:diff-form} the simple form 
\begin{equation}
  \label{eq:newscal}
 \la \xi,\eta\ra_H 
= -\frac{1}{2} \int_\R \int_\R \oline{\xi'(x)}\eta'(y) |x-y|^{2H}\, dx\, dy. 
\end{equation}

\subsection{Unitarity of the representations $U^H$, $0 < H < 1$} 

To verify the unitarity of the representations $U^H$, for 
$H > \shalf$, we calculate 
for $\xi,\eta \in \cS(\R)$
\begin{align*}
\frac{\la U^H_g \xi,   U^H_g \eta \ra}{H(2H-1)} 
&= \int_{\R} \int_{\R}  \oline{\xi(g^{-1}.x)}\eta(g^{-1}.y) 
\frac{|ad-bc|^{2H}}{|cx + d|^{2H} |cy + d|^{2H}} \frac{dx\ dy}{|x-y|^{2-2H}} \\ 
&= \int_{\R} \int_{\R}  \oline{\xi(x)}\eta(y) 
\frac{|ad-bc|^{2H-2}}{|c(g.x) + d|^{2H-2} |c(g.y) + d|^{2H-2}}
 \frac{dx\ dy}{|g.x-g.y|^{2-2H}}, 
\end{align*}
so that unitarity follows from 
\begin{align*}
c\cdot (g.x) + d = \frac{ad-bc}{a-cx}, 
\quad \mbox{ which implies} \quad 
\frac{|g.x-g.y|}{|x-y|}
& 
= \frac{|ad-bc|}{|a-cx| |a-cy|} 
= \frac{|c(g.x) + d|\cdot |c(g.y) + d|}{|ad-bc|}.
\end{align*}

For $H< \shalf$ we use the fact that $\PGL_2(\R)$ is generated by 
the affine group $\Aff(\R)$ and the map $\sigma(x) = x^{-1}$. 
As the kernel function $|x-y|^{2H}$ is translation invariant, 
the translations define unitary operators 
$U^H_g(\xi)(x) := \xi(x + b)$, $b \in \R$. 
For dilations $g^{-1}(x) = ax$, we have 
\[ (U^H_g\xi)(x) = \sgn(a) |a|^H \xi(ax) \quad \mbox{ and } \quad  
(U^H_g\xi)'(x) = |a|^{H+1} \xi'(ax).\] 
This leads to 
\[\la U^H_g\xi, U^H_g\eta \ra 
= \int_\R \int_\R |a|^{2H+2} \oline{\xi'(ax)} \eta'(ay)|x-y|^{2H}\, dx\, dy 
= \int_\R \int_\R \oline{\xi'(x)} \eta'(y)|x-y|^{2H}\, dx\, dy 
= \la \xi, \eta \ra.\] 
The most tricky part is to verify that the 
operator $(U\xi)(x) := -|x|^{-2H} \xi(x^{-1})$ is unitary on $\cH_H$. 
Since $U$ is an involution, it suffices to show that 
\begin{equation}
  \label{eq:sesq}
\la U\xi, \eta \ra   = \la \xi, U \eta \ra \quad \mbox{ for } \quad 
\xi,\eta \in \cS(\R). 
\end{equation}
To verify this symmetry relation, we may assume that $\xi$ and $\eta$ are real-valued. 
We first observe that 
\[ (U\xi)'(x) = 2 H \sgn(x)|x|^{-2H-1} \xi(x^{-1}) 
+ |x|^{-2H-2} \xi'(x^{-1}).\] 
This leads to 
\begin{align}\label{eq:39}
\la U\xi, \eta \ra 
&= -\int_\R \int_\R H \sgn(x)|x|^{-2H-1} \xi(x^{-1})\eta'(y)|x-y|^{2H}\, dx\, dy \\
&\ \ \ - \frac{1}{2} \int_\R \int_\R |x|^{-2H-2} \xi'(x^{-1})\eta'(y)|x-y|^{2H}\, dx\, dy.
\notag\end{align}
The second summand in \eqref{eq:39} equals 
\[ - \frac{1}{2} \int_\R \int_\R |x|^{2H+2} \xi'(x)\eta'(y)|x^{-1}-y|^{2H}
\, \frac{dx}{x^2}\, dy 
= - \frac{1}{2}  \int_\R \int_\R  \xi'(x)\eta'(y)|1-xy|^{2H}| \, dx\, dy,\] 
which is symmetric in $\xi$ and $\eta$.
The first summand in \eqref{eq:39} equals 
\begin{align*}
& -\int_\R \int_\R H \sgn(x)|x|^{2H+1} \xi(x)\eta'(y)|x^{-1}-y|^{2H}\, \frac{dx}{x^2}\, dy
= -\int_\R \int_\R H  \xi(x)\eta'(y)|1-xy|^{2H}\, dx\, \frac{dy}{x}.
\end{align*}
With 
\begin{align*}
& \int_\R \eta'(y)|1-xy|^{2H}\, \frac{dy}{x} 
= - 2H \int_\R \eta(y)|1-xy|^{2H-1}\sgn(1-xy)(-x)\, \frac{dy}{x} \\
&=  2H \int_\R \eta(y)|1-xy|^{2H-1}\sgn(1-xy)\, dy, 
\end{align*}
we obtain for the first summand in \eqref{eq:39} 
\[  -2H^2\int_\R \int_\R \xi(x)\eta(y)
|1-xy|^{2H-1}\sgn(1-xy)\, dx\, dy.\]
Again, the symmetry of the integral kernel now implies that 
this expression is symmetric in $\xi$ and~$\eta$.
We conclude that $U$ defines a unitary operator on $\cH_H$. 
This implies unitarity of the operators $U^H_g$ for $0 < H < \shalf$.

\section{The spectral measure of fractional Brownian motion} 
\mlabel{app:e}

For $\Re z > 0$, we have the integral representation of the Gamma function 
\begin{equation}
  \label{eq:gamma}
  \Gamma(z) = \int_0^\infty e^{-\lambda} \lambda^{z-1}\, d\lambda.
\end{equation}
For $0 < \alpha < 1$, this leads to 
\begin{equation}
  \label{eq:gamma2}
 \Gamma(1-\alpha) = \int_0^\infty e^{-\lambda} \frac{d\lambda}{\lambda^\alpha}.  
\end{equation}
By partial integration, we further obtain 
\begin{equation}
  \label{eq:gamma3}
\int_0^\infty (1 - e^{-\lambda})\lambda^{-1-\alpha}\, d\lambda 
= \frac{1}{\alpha} \int_0^\infty e^{-\lambda}\frac{d\lambda}{\lambda^\alpha} 
= \frac{\Gamma(1-\alpha)}{\alpha}.
\end{equation}
This in turn leads to 
\begin{equation}
  \label{eq:gamma4}
z^\alpha = \frac{\alpha}{\Gamma(1-\alpha)} 
\int_0^\infty (1 - e^{-z\lambda})\lambda^{-1-\alpha}\, d\lambda \quad \mbox{ for } \quad 
0 < \alpha < 1, z \in \C \setminus (-\infty,0] 
\end{equation}
(cf.\ \cite[p.~78]{BCR84}). 
In fact, for $z > 0$ real, this follows from \eqref{eq:gamma3} by dilation, so that 
the claim follows by analytic continuation from the holomorphy of both sides. 
For $z = i$, we obtain from \eqref{eq:gamma4} 
\begin{equation}
  \label{eq:gamma5}
\cos\Big(\frac{\alpha\pi}{2}\Big) = \Re e^{i\alpha\pi/2} 
= \Re(i^\alpha) = \frac{\alpha}{\Gamma(1-\alpha)} 
\int_0^\infty (1 - \cos \lambda)\lambda^{-1-\alpha}\, d\lambda \quad \mbox{ for } \quad 
0 < \alpha < 1. 
\end{equation}
With 
\begin{equation}
  \label{eq:gamma6}
  \Gamma(1-\alpha)\Gamma(\alpha) 
= \frac{\pi}{\sin(\pi \alpha)} 
= \frac{\pi}{2 \sin(\pi\alpha/2)\cos(\pi \alpha/2)}
\end{equation}
we obtain 
\begin{equation}
  \label{eq:gamma7}
  2\Gamma(1-\alpha)\cos\Big(\frac{\pi\alpha}{2}\Big) 
= \frac{\pi}{\sin\big(\frac{\pi\alpha}{2}\big)\Gamma(\alpha)}.
\end{equation}
Thus 
\[ \int_\R (1-\cos \lambda)|\lambda|^{-1-\alpha}\, d\lambda 
= \frac{\pi}{\alpha\Gamma(\alpha)\sin\big(\frac{\alpha\pi}{2}\big)}
= \frac{\pi}{\Gamma(1+\alpha)\sin\big(\frac{\alpha\pi}{2}\big)}
= \frac{\Gamma\big(1-\frac{\alpha}{2}\big)\Gamma\big(\frac{\alpha}{2}\big)}
{\Gamma(1+\alpha)}.\] 
We further obtain for $t \in \R$ by dilation 
\[ \int_\R (1-\cos t\lambda)|\lambda|^{-1-\alpha}\, d\lambda 
= \frac{\Gamma\big(1-\frac{\alpha}{2}\big)\Gamma\big(\frac{\alpha}{2}\big)}
{\Gamma(1+\alpha)}|t|^\alpha.\] 
For $H = 2 \alpha$ this implies 
\begin{align*}
|t|^{2H} 
&=\frac{\Gamma(2H+1)}{\Gamma(H)\Gamma(1-H)} 
\int_\R \frac{1-\cos \lambda t}{|\lambda|^{2H}}\, \frac{d\lambda}{|\lambda|} \\
&= \frac{1}{2}\frac{\Gamma(2H+1)}{\Gamma(H)\Gamma(1-H)} 
\int_\R \frac{(1-\cos \lambda t)^2 + \sin^2 \lambda t}{\lambda^2} 
|\lambda|^{1-2H}\, d\lambda \\
&= \frac{1}{2}\frac{\Gamma(2H+1)}{\Gamma(H)\Gamma(1-H)} 
\int_\R |e_t(\lambda)|^2 |\lambda|^{1-2H}\, d\lambda,  
\end{align*}
and therefore 
the spectral measure of fractional Brownian motion is 
\begin{equation}
  \label{eq:e.1}
d\sigma(\lambda) 
= \frac{1}{2}\frac{\Gamma(2H+1)}{\Gamma(H)\Gamma(1-H)} 
\cdot |\lambda|^{1-2H}\, d\lambda 
= \frac{\sin(\pi H)\Gamma(2H+1)}{2\pi}
\cdot |\lambda|^{1-2H}\, d\lambda.
\end{equation}

\end{document}